%% file: main.tex
\documentclass[12pt,preprint,nonatbib]{elsarticle}

\usepackage{geometry}
\usepackage{mathptmx}

\usepackage{setspace}
\usepackage{amsmath}
\usepackage{amssymb}

\usepackage{booktabs}
\usepackage{threeparttable}

\usepackage{graphicx}
\graphicspath{{./}}

\usepackage[section]{placeins}
\usepackage{float}

\usepackage{caption}
\usepackage[style=apa,backend=biber,hyperref=true]{biblatex}
\usepackage{xcolor}

\PassOptionsToPackage{hyphens}{url}

\usepackage{hyperref}
\hypersetup{
    colorlinks=true,
    linkcolor=blue,
    citecolor=blue,
    urlcolor=blue,
    pdftitle={Synthetic Personalities: How Well Can LLMs Mimic Individual Respondents Using Socio-Economic Microdata?},
    pdfauthor={Leonard Kinzinger, Jochen Hartmann}
}

\begin{document}

\input{00_titlepage}

\newpage

\setlength{\parindent}{0pt}
\setlength{\parskip}{1em}

\input{01_introduction}
\input{02_literature_review}
\input{03_methodology}
\input{04_results}
\input{05_discussion}

\printbibliography

\end{document}

%% file: 00_titlepage.tex
\begin{frontmatter}

\title{Synthetic Personalities: How Well Can LLMs Mimic Individual Respondents Using Socio-Economic Microdata?}

\author[a]{Leonard Kinzinger\corref{cor1}}
\ead{leonard.kinzinger@tum.de}
\author[a]{Jochen Hartmann}
\address[a]{TUM School of Management, Technical University of Munich, 80333 Munich, Germany}
\cortext[cor1]{Corresponding author}

\begin{abstract}
LLM-based digital twins promise to scale and accelerate market research, but most published twins are either coarse persona bots conditioned on a few demographic questions or detailed individual-level twins built on purpose-collected surveys and interview transcripts. Neither setup speaks to the operationally most relevant case for marketing practice: building detailed individual twins from the pre-existing heterogeneous panel data that firms already accumulate through CRM systems, loyalty programs, and repeat surveys. We construct detailed individual-level twins from the German Socio-Economic Panel (SOEP) and evaluate them across a $3 \times 5 \times 2 \times 2$ construction-method grid that covers three open-weights LLMs, five cumulative information depths ranked by normalized Shannon entropy, two embedding methods, and two reasoning modes, scoring over 2.1 million twin responses on 500 participants and 183 held-out questions. Twin quality rises with information depth but with diminishing returns past the 75 percent entropy quartile, which acts as a cost-efficient Pareto point relative to the best-performing 100 percent cells. Switching the embedding from a narrative persona summary to a raw dialog history of past responses raises hold-out accuracy in every model-by-reasoning cell at the 100 percent depth, while an explicit thinking mode raises rank-order correlation without moving accuracy. Best-cell accuracy reaches 78.8 percent and Fisher-$z$ correlation reaches $r = 0.590$ on the SOEP held-out evaluation set. The findings suggest that twin-based market research is no longer gated by data design, but by item volume, model selection, and a small set of construction-level decisions that this paper now maps.
\end{abstract}


\end{frontmatter}

%% file: 01_introduction.tex
\section{Introduction}
\label{sec:introduction}

Traditional market research, including large-sample surveys, conjoint experiments, and in-depth interviews, is invaluable for understanding consumers. But it is also slow, expensive, and especially difficult for hard-to-reach respondents such as low-incidence demographics and senior decision-makers \parencite{arora2025ai}. Large language models (LLMs) as synthetic respondents promise a step change in cost and speed. Marketing research has shown that LLMs can estimate within-category willingness-to-pay after fine-tuning on category data \parencite{brand2023using}, reproduce brand-perception maps that align with revealed-preference data \parencite{li2024frontiers}, recover individual-level time preferences \parencite{goli2024frontiers}, cut the survey data needed for conjoint estimation by 25 to 80 percent \parencite{wang2026large}, and calibrate LLM-generated samples to reduce their distributional disparity from human responses \parencite{leng2024reduce}. Commercial vendors such as simile.ai are forming around these capabilities and now sell on-demand synthetic respondent panels.\footnote{simile.ai closed a USD~100~million funding round in February~2026 to help firms predict human behavior at scale (see \url{https://www.bloomberg.com/news/articles/2026-02-12/ai-startup-nabs-100-million-to-help-firms-predict-human-behavio}), an indication that synthetic respondents are moving from research curiosity to a market-relevant product.}

The synthetic-respondent literature has organized itself around two granularities. Coarse representations condition an LLM on a handful of demographic attributes and represent a segment of people in aggregate \parencite{argyle2023outofone,santurkar2023whose,brand2023using,li2024frontiers,wang2026large,goli2024frontiers}. Detailed individual representations, in contrast, condition the model on rich individual-level data and represent one specific real person \parencite{park2026llmagentsgroundedselfreports,Toubia2025,peng2025digital}. The latter is strictly more flexible because individual-level twins can be aggregated post hoc to any segment a researcher cares to define, while coarse persona bots cannot be decomposed back to individuals. Academic work in the detailed paradigm has, however, so far been built almost exclusively on datasets collected for the purpose of creating digital twins. \textcite{Toubia2025} build the Twin-2K-500 dataset as a four-wave instrument designed end-to-end for twin construction with more than 500 items per respondent. \textcite{park2026llmagentsgroundedselfreports} run two-hour AI-conducted semi-structured interviews per respondent. \textcite{binz2025foundation} and \textcite{binz2026post} curate the Psych-101 and Psych-201 trial-by-trial behavioral batteries across more than 60{,}000 participants and use them to fine-tune and benchmark the Centaur family of cognition models. This is not the data shape firms hold. Enterprise CRM systems, loyalty databases, and repeat-survey programs accumulate heterogeneous item banks over years without a twin-oriented design, with uneven information density per item. Less is known about whether useful individual-level twins can be built from the kind of panel data that already sits on a firm's disk. 

A second open question concerns construction methodology: while embedding method, reasoning mode, and information depth have each been examined anecdotally \parencite{park2026llmagentsgroundedselfreports,Toubia2025,peng2025digital}, no study to our knowledge has compared them systematically on a shared dataset and a shared evaluation set.

We address both gaps. We construct individual-level digital twins from the German Socio-Economic Panel (SOEP), a 41-year longitudinal panel that interviewed more than 28{,}000 respondents in 2023 and exposes more than 900 individual-level question-answer pairs per respondent in the 2023 wave alone \parencite{goebel_german_2019,soep_v40eu_2025}. The SOEP serves as an academic analogue of the heterogeneous panel data that firms accumulate through CRM systems, loyalty programs, and repeat surveys. We then evaluate twins across the full $3 \times 5 \times 2 \times 2$ construction-method grid (three open-weights LLMs, five cumulative information depths ranked by normalized Shannon entropy, two embedding methods, two reasoning modes), scoring over 2.1 million individual twin responses on 500 participants and 183 held-out questions against ground truth, using the accuracy, rank-order correlation, and dispersion-ratio framework established by \textcite{peng2025digital}, \textcite{Toubia2025}, and \textcite{park2026llmagentsgroundedselfreports}. Intuitively, accuracy measures how close each twin's answer sits to its human counterpart's answer on the question's natural scale (1 for a perfect match, 0 for the maximum possible miss; for example, a twin who answers 6 to a human's 5 on a 1-to-7 scale scores 0.833), while rank-order correlation measures whether the twins reproduce who scores higher than whom across participants on each question, i.e., the individual-level heterogeneity that humans display \parencite{peng2025digital}. We use open-weights models because they are reproducible end-to-end and remain deployable in license-bound or regulated settings where third-party APIs cannot be used.

Several patterns emerge from the construction-method grid. Accuracy and rank-order correlation both rise monotonically with information depth on a concave curve: the 75 percent quartile of entropy-ranked items captures the bulk of the available signal at a moderate prompt-token cost, while the 100 percent quartile attains the best cells at substantially higher cost. Switching the embedding from a narrative persona summary to a raw dialog history of past survey responses raises accuracy at the 100 percent depth for every one of the three models, in both reasoning modes, while an explicit thinking mode raises rank-order correlation without moving accuracy.
Across the three open-weights models, Qwen 3 \parencite{qwen3_2025} leads on average across all three metrics, with Gemma 4 \parencite{gemma4_2025} attaining the single best accuracy cell at 78.8 percent and Qwen 3 the highest rank-order correlation ($r = 0.590$), both in the dialog-thinking-100-percent cell. On the SOEP held-out evaluation set, these levels sit at the same order of magnitude as the leading published twins built on purpose-collected data \parencite{Toubia2025,park2026llmagentsgroundedselfreports,peng2025digital}.

Our contribution centers on demonstrating that individual-level twins can be built from pre-existing heterogeneous panel data of the kind firms already accumulate, opening the digital twin program beyond the bespoke surveys and interview protocols used to date. The construction-method grid we deliver on a shared dataset and a shared evaluation set identifies which choice drives which metric and at what cost, and locates a clear Pareto point at the 75 percent entropy quartile with the maximum at 100 percent, giving practitioners an actionable rule for context budgeting.

\FloatBarrier

%% file: 02_literature_review.tex
\section{Background and Related Literature}
\label{sec:background}

Our paper draws on three lines of work: LLMs as synthetic survey respondents in marketing, individual-level digital twins built from rich human data, and the diagnostic literature on their failure modes.

\subsection{LLMs as Synthetic Survey Respondents in Marketing}
\label{sec:llms_in_marketing}
Marketing research has shown that LLMs can act as synthetic respondents on several aggregate tasks, and the discipline has been framing this opportunity since at least \textcite{peres2023chatgpt} and \textcite{arora2025ai}. \textcite{yoo2025whole} extend this framing into a field guide that replicates 35 consumer-research articles across all six research stages on ChatGPT-4o. \textcite{brand2023using} show that off-the-shelf GPT estimates of willingness-to-pay are often inaccurate or wrong-signed, while fine-tuning on category-specific human conjoint data recovers the sign and rough magnitude of within-category attribute valuations. \textcite{li2024frontiers} validate off-the-shelf GPT-4 brand-perception maps against human surveys and revealed-preference car-trade-in data, reaching over 75 percent agreement on car-brand similarity ratings. Both papers report a clear ceiling on demographic heterogeneity: Brand et al.\ find that GPT cannot meaningfully reflect heterogeneous \emph{preferences} across demographic groups, and Li et al.\ find that while LLM-generated brand \emph{perceptions} are largely invariant across demographics (mirroring human data), brand \emph{preferences} can be aligned only when the prompt is explicitly conditioned on individual attributes. \textcite{wang2026large} pool LLM-generated and real choice data in a transfer-learning estimator with proved asymptotic consistency and recover efficiency gains of 25 to 80 percent on two conjoint studies. \textcite{goli2024frontiers} extend LLM-based prediction to individual-level time preferences. \textcite{leng2024reduce} propose an optimal sample-calibration scheme that reweights LLM-generated responses to reduce their distributional disparity from human samples. Adjacent work uses LLMs for feature extraction in high-stakes hiring contexts \parencite{chakraborty2025can}.

Response stability across this work is uneven. \textcite{sarstedt2024using} document substantial drift across model versions and prompt forms, and \textcite{brucks2025prompt} show in 5,760 GPT-4 factorial-experiment responses a 63 percent first-position selection rate and a 74 percent rate of choosing B over C under letter labels, with the first-position bias falling to 54 percent when the same options carry symbol rather than letter labels. \textcite{Toubia2025} reframe the question by benchmarking twin responses against the underlying human two-week test-retest accuracy on Twin-2K-500, positioning the human self-consistency ceiling as the natural reference point for twin stability rather than absolute accuracy. Industry benchmarks point in the same direction, with LLM accuracy degrading as the target segment narrows \parencite{stromberg2025blind}.

\FloatBarrier

\subsection{Two Paradigms of Synthetic Respondents}
\label{sec:two_paradigms}

Synthetic-respondent work has organized itself around two granularities, which we term \emph{coarse representations} and \emph{detailed individual representations}. Coarse representations condition on a few demographics (age, gender, race, party) and are most often used to represent a segment of people \parencite{argyle2023outofone,santurkar2023whose,aher2023using,sarstedt2024using}. Its output is by construction an aggregation, and it is evaluated against aggregate human distributions: opinion-distribution match \parencite{santurkar2023whose,suh2025language}, mean treatment-effect replication \parencite{hewitt2024predicting,aher2023using}, and between-group distributional fidelity \parencite{boelaert2025machine}. Detailed individual representations condition an LLM on rich individual-level data and represent one specific real person \parencite{park2026llmagentsgroundedselfreports,Toubia2025}. 
The unit of analysis is finer than a segment, and the same twins can be aggregated, post hoc, to any segment the researcher cares to define and to alternative segment definitions in parallel. The aggregation direction is asymmetric: persona bots cannot be decomposed into individuals, but twins can be summed up to any segment a researcher constructs. 
Evaluation therefore operates at both levels. At the individual level, the standard metrics are accuracy and rank-order correlation \parencite{Toubia2025,park2026llmagentsgroundedselfreports,peng2025digital}, together with the variance ratio of twin-to-human response dispersion \parencite{peng2025digital}, each measured per question across respondents; at the population level, twins inherit the persona-bot metrics above. \textcite{anthis2025position} and \textcite{kozlowski2025simulating} give higher-level framings of this space. We use the term \emph{digital twin} in the narrow individual sense throughout.

\FloatBarrier

\subsection{Building Individual-Level Twins}
\label{sec:building_twins}

A digital twin is built from the real responses of one individual, and its task is to answer new questions as that person would. Inputs vary along several dimensions: the type of question (closed-scale, free-form, behavioral task), the response modality (typed survey, conversation transcript, voice file), the level of detail (a handful of demographics versus hundreds of items per person), and the topics covered. No input set ever fully captures a person, meaning every digital twin is by design a lossy representation. The goal of a digital twin is not to represent a human to 100\%, but to be close enough that their outputs are realistic (reproducing the human's specific knowledge, opinions, and characteristic gaps, including bounded-rationality patterns documented in behavioral economics) to be useful. This means that the behavior is such that it could have been observed by a similar person, and that digital twins and real humans are close enough that they share similar aggregated metrics.

Published twins differ mainly in the type of input data and how it is fed to the model. \textcite{park2026llmagentsgroundedselfreports} use two-hour AI-conducted semi-structured interview transcripts (about 6,500 words per person), injected directly into the prompt, for a representative US sample of 1,052 respondents. \textcite{park2026llmagentsgroundedselfreports} normalize their accuracy by each participant's two-week test-retest consistency: the agent's prediction accuracy is divided by the participant's own replication accuracy on the same items, so a normalized accuracy of 1.0 corresponds to the agent matching the participant's own self-consistency two weeks later. They report normalized accuracy of 0.83 (interview-only) and 0.86 (interview-plus-survey) on the General Social Survey, against 0.74 for a demographics-only baseline, 0.80 on the Big Five inventory, and 0.66 on economic games. \textcite{Toubia2025} take the structured-survey route with the Twin-2K-500 dataset: a four-wave survey of 2,058 US respondents covering 14 demographic items, 19 personality tests (279 questions, 26 constructs), 11 cognitive measures (85 questions), 10 economic-preference tests (34 questions), 21 behavioral-economics paradigms (48 questions), and a 40-question pricing study, with individual-level accuracy near 0.72 absolute and 0.88 normalized against participant test-retest. Both lines of work require data collected for the purpose of twin-building.

A parallel research stream fine-tunes models on large collections of human response data rather than feeding individual histories at prompt time \parencite{kolluri2025finetuning,suh2025language,orlikowski2025beyond}. The most thorough examples are \textcite{binz2025foundation} and \textcite{kolluri2025finetuning}. \textcite{binz2025foundation} fine-tune LlaMA-3.1 70B on the Psych-101 dataset with trial-by-trial data from more than 60,000 participants and 10,000,000 choices across 160 experiments. Their Centaur model not only improves performance on held-out participants but also generalizes to new tasks and domains. \textcite{binz2026post} extend this line with Psych-201, a behavioral-alignment benchmark used to show that post-training systematically reduces alignment with human behavior across model families and that persona-induction on top of an off-the-shelf chat model does not improve individual-level predictions on that battery. \textcite{kolluri2025finetuning} fine-tune LLaMA-3-8B and Qwen-2.5-14B on 2.9 million responses across 210 NSF-TESS social-science experiments and report distributional alignment with human responses that beats GPT-4o by 12.1 percent (LLaMA) and 13.2 percent (Qwen) on completely unseen studies. These approaches are complementary to prompt-based methods but require curated training data and a separate compute budget.

What is currently missing in published work are digital twins, built from pre-existing heterogeneous panel datasets, of the kind market research firms already hold or that companies routinely collect through CRM, loyalty, or repeat-survey programs. Such panels are a strong candidate substrate for twin construction for at least three reasons. First, they accumulate hundreds of items per person through repeated surveys across many topics, which is materially more than the demographics-only personas in published persona-bot work and on par with or larger than purpose-built twin instruments. Second, the topics covered are commercially relevant (consumption, employment, family, household equipment, time use), so twins built on them speak directly to the questions firms ask in practice. Third, panels run for years, so values, opinions, and behaviors are observed repeatedly per person and life-course changes are recorded, which gives the model genuine within-person variation rather than a single cross-sectional snapshot. 

\FloatBarrier

\subsection{Documented Failure Modes}
\label{sec:failure_modes}

At the segment level, prior work has shown that persona bots produce opinion distributions that systematically deviate from human aggregates \parencite{santurkar2023whose,boelaert2025machine,bisbee2024synthetic} and can carry the political leanings introduced through pretraining and alignment \parencite{santurkar2023whose,motoki2024more,rozado2024political}. At the individual level, \textcite{peng2025digital} reports that twins inherit these biases and add their own, including shrinkage toward the base-model prior and over-reliance on demographic shortcuts, and \textcite{li2026llm} finds that bias is amplified as more LLM-generated persona content is added; \textcite{park2024diminished} and \textcite{gupta2024bias} document related patterns of reduced diversity and persona-assigned reasoning bias. Because twins aggregate up into segments, individual-level failures propagate to the population level whenever segments are reconstructed from twin populations.

\textcite{peng2025digital} report a 19-substudy pre-registered evaluation using the Twin-2K-500 dataset. Across 164 outcomes, individual-level accuracy is 0.748 for the full persona and 0.746 for a 14-variable demographics-only persona, statistically indistinguishable at p = 0.37, with an empty persona reaching 0.734 and a random null 0.629. Per-outcome cross-respondent correlation is r = 0.197 for the full persona. Twin response variance is lower than human variance in 154 of 164 outcomes (94 percent). The authors consolidate these patterns into five distortions: insufficient individuation, demographic stereotyping, representation bias, ideological drift, and hyper-rationality.

Companion diagnostic work documents related concerns from adjacent angles, including psychometric invalidity of LLM responses on standardized scales \parencite{petrov2024limited}, prompt-form and probe sensitivity \parencite{dominguezolmedo2024questioning,wang2024myanswer,brucks2025prompt}, alignment-induced ideological drift \parencite{lyman2025balancing}, social-desirability response bias \parencite{salecha2024large}, and contamination risk for survey research at the population level \parencite{westwood2025potential,gao2025take}. A parallel literature uses LLMs to replicate prior human-subject studies and treatment effects, with mixed but informative results \parencite{cui2025large,mei2024turing,aher2023using,hewitt2024predicting}.

We use the insights from this literature to interrogate the outputs of our own digital twins, reporting the individual- and population-level versions of the metrics that this stream has established and noting where our results align with or diverge from the biases cataloged above.

\FloatBarrier

\subsection{The Gap We Fill}
\label{sec:gap}

Our contribution sits at the intersection of three streams of work that together do not address the operationally most relevant question for marketing practice. We work in the detailed individual-representation paradigm rather than the coarse persona-bot paradigm: most LLM market-research work to date \parencite{brand2023using,li2024frontiers,goli2024frontiers,wang2026large} conditions on a handful of demographic axes and aggregates LLM outputs to population-level targets, which cannot speak to within-person fidelity, rank-order, or dispersion and cannot be decomposed back to individuals. The detailed twin literature \parencite{Toubia2025,park2026llmagentsgroundedselfreports,binz2025foundation} is individual-aware but bespoke-data-only, requiring data collection designed end-to-end for twin construction, which is precisely what firms cannot retrofit cheaply. Construction-method levers such as embedding, reasoning, and information depth have each been examined anecdotally \parencite{park2026llmagentsgroundedselfreports,Toubia2025,peng2025digital,wei2022chain,snell2024scaling} but never systematically on a shared dataset and a shared evaluation set. We fill this intersection by building detailed individual-level twins from a pre-existing heterogeneous panel dataset (SOEP) and evaluating them across the full construction-method grid using the accuracy, rank-order correlation, and dispersion-ratio framework inherited from \textcite{peng2025digital}, \textcite{Toubia2025}, and \textcite{park2026llmagentsgroundedselfreports}.

\FloatBarrier

%% file: 03_methodology.tex
\section{Methodology}
\label{sec:methodology}

Prior work has shown that LLM-based digital twins are sensitive to three independent construction choices: how persona context is embedded in the prompt \parencite{park2026llmagentsgroundedselfreports,Toubia2025}; whether the model is prompted to reason explicitly before responding \parencite{Toubia2025}; and how much individual context the prompt carries \parencite{park2026llmagentsgroundedselfreports,peng2025digital}. However, to the best of our knowledge, no published study has compared digital-twin construction methods systematically, which leaves an open question: which combination of construction choices actually performs best?

The remainder of this section sets out the design of the resulting $3 \times 5 \times 2 \times 2 = 60$-cell construction-method grid: the SOEP source data and our 500-participant sample (Section~\ref{sec:data_sample}), the three open-weights LLMs (Section~\ref{sec:llm_selection}), the two embedding methods, the two reasoning modes, the five cumulative information depths ranked by Shannon entropy (Section~\ref{sec:information_depth}), and the three evaluation metrics (Section~\ref{sec:phase1_metrics}). By holding the participant subsample and the held-out evaluation set fixed across all 60 cells, the resulting curves isolate the marginal contribution of each construction-method choice and let us assess its robustness across models.

\FloatBarrier

\subsection{Data and Sample Selection}
\label{sec:data_sample}

We build our digital twins on the German Socio-Economic Panel (SOEP), a longitudinal household panel administered by DIW Berlin since 1984 \parencite{goebel_german_2019} that provides individual-level microdata on a representative cross-section of the resident German population. It covers a heterogeneous adult population across ten thematic life domains and spans four decades of repeated measurement; we accessed the microdata under a granted SOEP data contract with DIW Berlin. 
As a non-U.S. panel, it also serves as a counter-weight to the U.S.-centric datasets that dominate published digital twin work \parencite{Toubia2025, park2026llmagentsgroundedselfreports}. Specifically, we use SOEP-Core v40 EU Edition \parencite{soep_v40eu_2025}, which covers Demographics and Population; Integration, Migration, and Transnationalisation; Health and Care; Employment and Occupation; Education and Qualification; Family and Social Networks; Attitudes, Values, and Personality; Housing, Equipment, and Private Household Services; Time Use and Environmental Behaviour; and Income, Taxes, and Social Security. Table~\ref{tab:dataset_overview} reports the resulting question composition across these ten domains.

We take a cross-sectional snapshot of the 2023 wave and merge five source datasets: \texttt{pl} (individual questionnaire), \texttt{biol} (biography), \texttt{hl} (household), \texttt{ppathl} (person meta-data), and \texttt{pgen} (generated status variables). We retain all migration samples \parencite{brucker_new_2014} and refugee samples \parencite{brucker_kosyakova_2025} to preserve the full diversity of the panel.\footnote{Recent SOEP recruitment waves have disproportionately sampled refugees and migrants (particularly from Ukraine and Syria).}

\input{tab_dataset_overview}
Starting from 820{,}360 individual records in the \texttt{pl} dataset, we apply participant-level filters (2023 survey wave, complete Big Five responses on the BFI-S \parencite{schupp_big_2008}, biography coverage within 2019 to 2023, valid household linkage) and a column-level pipeline (versioned-column deduplication, fill-rate threshold, technical-identifier exclusion, near-duplicate removal, and LLM-based question reformulation and scale-type assignment) to arrive at 16{,}055 participants and 949 usable question-answer pairs per participant across all ten SOEP categories. The 949 questions split into 38 core demographic items that always appear in the persona context, 728 persona-context questions used to build the digital twin, and 183 held-out evaluation questions reserved exclusively for measuring prediction accuracy. Step-by-step row counts, filter rationales, and LLM post-processing examples are reported in Web Appendices A.3, A.4, and A.5.

For the construction-method grid we draw a fixed random subsample of 500 participants from the 16{,}055-participant pool. Five hundred participants is a deliberate sweet spot: it is large enough to make differences between cells in the 60-cell grid detectable with reasonable statistical power, small enough to keep the total compute requirement tractable (the grid runs over 2.1 million twin responses locally, each of which then has to be scored again locally by an LLM-as-judge), and small enough to avoid a setting in which trivially small construction-method differences become statistically significant by sheer sample size. Holding this 500-participant subsample fixed across all 60 cells is what makes the cell-to-cell comparisons interpretable.

Although SOEP is academic microdata, datasets of comparable shape (detailed individual-level survey histories spanning many life domains) also exist in large market-research firms, in enterprises with mature CRM systems and in-house consumer panels, and increasingly in digital-twin startups such as simile.ai and ListenLabs. Findings on SOEP therefore have direct parallels in the commercial settings where comparable but proprietary data are available.

\FloatBarrier

\subsection{Large Language Model Selection}
\label{sec:llm_selection}

We evaluate three open-weights models: \emph{Qwen3-30B-A3B} \parencite{qwen3_2025}, \emph{Gemma-4-26B-A4B} \parencite{gemma4_2025}, and \emph{Ministral-3-14B} \parencite{liu2026ministral}. In the Results chapter we refer to them as: Qwen~3, Gemma~4, and Ministral~3. We selected them because each ships paired reasoning and non-reasoning variants on Hugging Face, and all three rank among the top open-source models runnable on a single GPU as of May~2026 on LiveBench \parencite{livebench} and the Artificial Analysis Intelligence Index.\footnote{\url{https://artificialanalysis.ai/models/open-source/small}} We serve all inference locally on a single NVIDIA H100 via vLLM; closed-source frontier models are excluded by the SOEP data license. All three models support German natively and we use the original SOEP wording. Multilingual benchmarks consistently show small accuracy gaps relative to English, attributable to English-skewed pretraining and tokenizer optimization. Our reported accuracy and correlation levels can therefore be read as a lower bound for what equivalent twins would achieve on English-language panels in the US or UK, where the same models have stronger native support. 
\FloatBarrier

\subsection{Embedding methods}

The persona context can be presented to the LLM either as a \emph{narrative summary} that compresses each participant's survey responses into a dense description of the information, or as a \emph{dialog input} that preserves the raw question-and-answer pairs as alternating user/assistant chat turns \parencite{park2026llmagentsgroundedselfreports,Toubia2025}. We construct the narrative summaries using a Chain-of-Density summarization approach, adapted from \textcite{adams_sparse_2023}. Full prompts, per-category examples, and the Chain-of-Density summarization procedure are reported in Web Appendix B.

\subsection{Reasoning modes}

Each of the three models is queried either in \emph{non-reasoning} mode, where it provides a direct response to the question, or in \emph{reasoning} mode, in which it first produces an internal \texttt{<think>} block before emitting the visible response. For Qwen 3 and Ministral 3 we use the dedicated reasoning checkpoints released by the vendors (Qwen3-30B-A3B-Thinking-2507 and Ministral-3-14B-Reasoning-2512); Gemma 4 does not ship a separate reasoning checkpoint, so we elicit the reasoning trace from the same Gemma-4-26B-A4B model via the structured-thinking prompt template in Web Appendix B.3. The non-reasoning variants are Qwen3-30B-A3B-Instruct-2507, Ministral-3-14B-Instruct-2512, and the same Gemma-4-26B-A4B model with the reasoning instruction removed. The design is inspired by chain-of-thought prompting \parencite{wei2022chain} and the broader test-time-compute literature \parencite{snell2024scaling}; the per-model system prompts and an example reasoning trace are reported in Web Appendix B.3.

\subsection{Information Depth}
\label{sec:information_depth}
\label{sec:information_density}
Prior work on individual-level digital twins has typically been built from data collected specifically for the purpose: the four-wave Twin-2K-500 dataset of $\sim$500 curated questions \parencite{Toubia2025}, and the two-hour American Voices Project interview transcripts \parencite{park2026llmagentsgroundedselfreports}. Our dataset is different and, we argue, more representative of the conditions under which firms would actually deploy twins. SOEP accumulates hundreds of items per person through repeated surveys across many commercially relevant life domains (consumption, employment, family, time use, housing, attitudes). It was never designed for twin construction, the topical coverage is uneven, and the information density per item varies widely. This is also the data shape held by firms with mature CRM, loyalty, and repeat-survey programs, and it raises three substantial questions for prompt construction: (1) Should all available information be included, or does additional context eventually hurt rather than help? (2) What is the marginal gain of each additional unit of context, and at what point do returns diminish? (3) On what principle should items be ranked when only a subset of data is selected?

We address all three by structuring the persona context by information density, using normalized Shannon entropy \parencite{shannon_1948, cover_thomas_2006} as a measure of how strongly each survey question differentiates between individuals. For each of the 728 persona-context questions we compute the normalized Shannon entropy of the empirical response distribution across all 16{,}055 participants in the pre-filtered pool, treating this value as a proxy for the question's differentiating power.
For a discrete random variable $X$ with $k$ possible outcomes and probability mass $p(x_i)$, we compute the normalized Shannon entropy as the raw entropy divided by the maximum entropy attainable for the question's scale:

\begin{equation}
    H_{\text{norm}}(X) = \frac{-\sum_{i=1}^{k} p(x_i) \, \log_2 p(x_i)}{\log_2 k} \in [0, 1].
    \label{eq:normalized_entropy}
\end{equation}

By construction, $H_{\text{norm}} = 0$ when all respondents give the same answer (no differentiating power) and $H_{\text{norm}} = 1$ when responses are uniformly distributed across categories (maximum differentiating power). Intuitively, a question where 99\% of respondents answer ``Yes'' tells us almost nothing about how to distinguish a specific person, while a question whose responses are evenly distributed across five categories provides substantial differentiating information.

We use these normalized entropy scores to rank the 728 persona-context questions in descending order and split at the quartile boundaries into four equal-sized buckets of 182 questions each. We then define five \emph{cumulative} information depth levels: (1) Basic demographic information, (2) Demographics + 25\% Quartile, (3) Demographics + 50\% Quartile, (4) Demographics + 75\% Quartile, (5) All available persona-context data. The depth levels are cumulative, meaning digital twins at depth level 3 see all demographic, first quartile and second quartile questions, sorted by normalized Shannon entropy. This cumulative ordering isolates the marginal effect of each successive quartile on prediction accuracy.

\begin{figure}[htb]
    \centering
    \includegraphics[width=\textwidth]{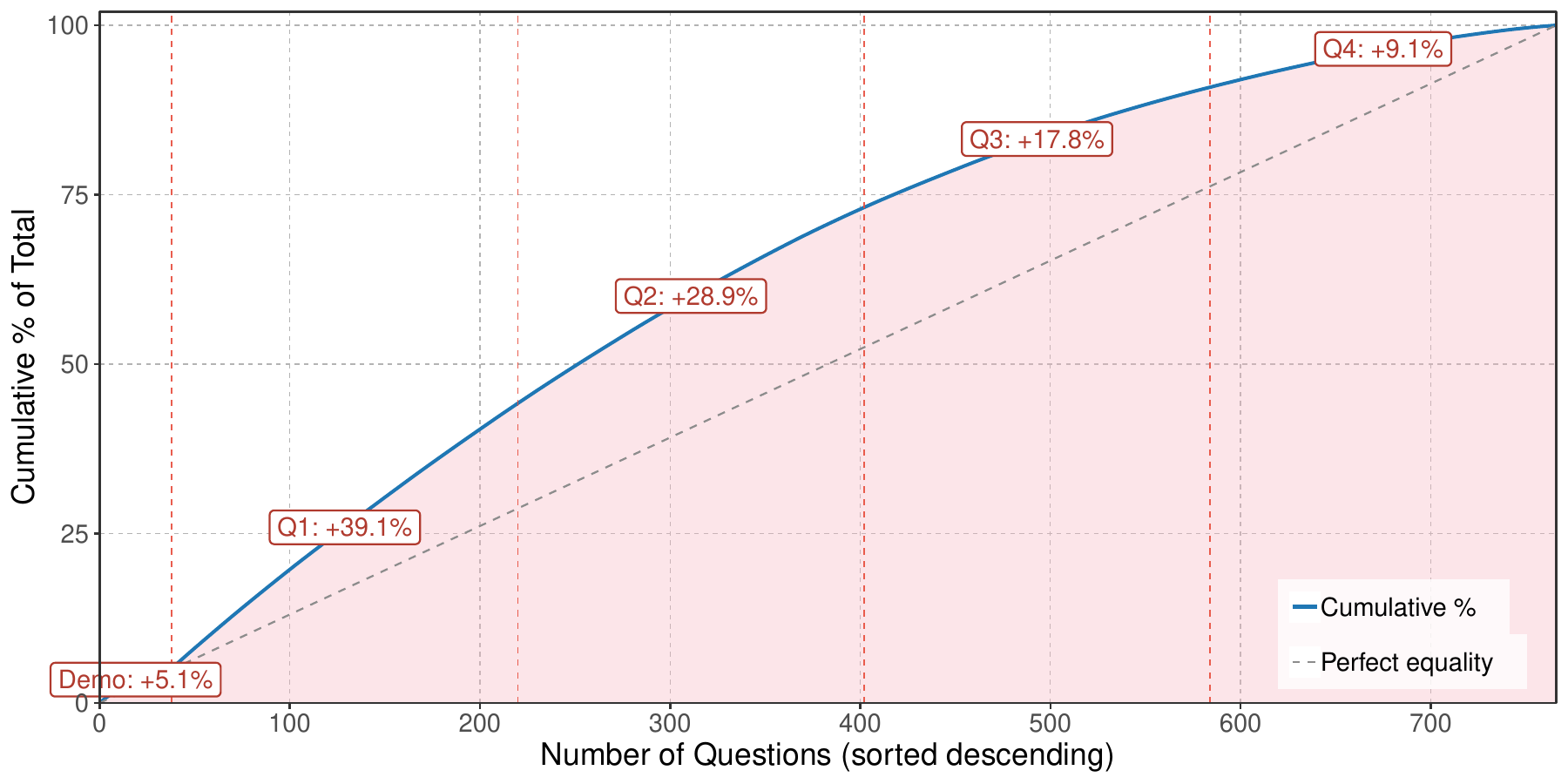}
    \caption{Cumulative distribution of normalized Shannon entropy across the 728 persona-context questions, sorted by descending entropy. Dashed vertical lines mark the boundaries of the four cumulative information depth levels. The dashed diagonal represents perfect equality (uniform entropy contribution).}
    \label{fig:entropy_cdf}
\end{figure}

The distribution of normalized entropy across the 728 persona-context questions is non-uniform (see cumulative distribution in Figure~\ref{fig:entropy_cdf}). The first entropy quartile alone captures 39.1\% of the total Shannon entropy across all persona-context questions, while the last entropy quartile contributes only 9.1\%. Our systematic comparison of construction methods reveals whether the accuracy of digital twins follows the same diminishing-returns shape. Web Appendix D.6 reports a random-question-selection ablation on a subsample of 500 participants, in which the four cumulative quartiles are replaced with cumulative random quartiles of matching size.

\FloatBarrier

\subsection{Evaluation Metrics}
\label{sec:phase1_metrics}

We evaluate twins on three complementary metrics, following \textcite{peng2025digital}. The first is an accuracy score that captures how close each twin's answer is to the human's. For each (question, participant) cell we adapt the score from \textcite{Toubia2025} and extend it with partial credit for ordinal items:
\begin{equation}
    S = \begin{cases}
        1 & \text{if prediction = ground truth (nominal)} \\
        0 & \text{if prediction $\neq$ ground truth (nominal)} \\[6pt]
        1 - \dfrac{|\text{pos}(\text{GT}) - \text{pos}(\text{Pred})|}{n_{\text{options}} - 1} & \text{(ordinal)} \\[6pt]
        1 - \dfrac{|\text{GT} - \text{Pred}|}{\text{Range}} & \text{(metric)}
    \end{cases}
    \label{eq:accuracy_score}
\end{equation}
where $\text{pos}(\cdot)$ is the position of an ordinal option and Range is the natural range of a metric outcome; $S \in [0,1]$ with 1 a perfect prediction. Predicted responses are normalized via an LLM-as-judge pass before scoring (see Web Appendix B.7).

The second is a correlation metric that captures whether twins additionally \emph{rank-order} participants the same way humans do, a property accuracy is blind to whenever twins sit close to the human mean on every question. For each question $q$ we compute the Pearson correlation $r_q$ between the vectors of human and twin responses across participants, restricted to ordinal and metric items where option order is meaningful. We aggregate the $r_q$ across questions via a Fisher $z$ transform (mean of $\operatorname{atanh}(r_q)$, back-transformed) to obtain the cell-level $\bar r$.

The third is a dispersion-fidelity metric that captures whether twins reproduce the \emph{spread} of human responses across participants, the failure mode \textcite{peng2025digital} describes as ``insufficient individuation.'' For each question $q$ we compute the ratio $\sigma^{(q)}_{\mathrm{twin}} / \sigma^{(q)}_{\mathrm{human}}$ of digital twin to human standard deviations across participants, on the same ordinal+metric items used for correlation, and report two summaries per cell: the mean ratio across questions (with a bootstrap CI) and, mirroring Peng's binary framing, the share of questions with $\sigma^{(q)}_{\mathrm{twin}} < \sigma^{(q)}_{\mathrm{human}}$. This variance ratio complements the other two metrics and tells us if the distribution of digital twins is too narrow or too wide. It is also essential to spot model-collapse and metric-scale blow-out failures.

For all three metrics we aggregate \emph{per question first}, then average across questions, matching the convention of \textcite{peng2025digital} and weighting every question equally regardless of participant coverage. Implementation details, edge-case conventions (zero-variance, twin-collapse, metric-scale winsorization), and the per-question coverage threshold are reported in Web Appendix C.

\FloatBarrier

%% file: tab_dataset_overview.tex
\begin{table}[htbp]
\caption{Dataset overview: distribution of 949 usable datapoints across SOEP thematic categories, data splits, and scale types. Values show column-wise percentages within each split.}
\label{tab:dataset_overview}
\centering
\small
\begin{tabular}{l r | rr r}
\toprule
\textbf{SOEP Category} & \textbf{\shortstack{Demographic\\Information}} & \textbf{\shortstack{Questions for\\Persona Constr.}} & \textbf{\shortstack{Questions for\\Persona Eval.}} & \textbf{Total} \\
 & ($n{=}38$) & ($n{=}728$) & ($n{=}183$) & ($N{=}949$) \\
\midrule
Employment and Occupation                 & 26.3\% & 14.3\% & 14.8\% & 141 \\
Education and Qualification               & 21.1\% &  7.4\% &  5.5\% &  72 \\
Demographics and Population               & 18.4\% &  7.1\% &  6.6\% &  71 \\
Income, Taxes, and Social Security        &  5.3\% & 10.3\% & 12.0\% &  99 \\
Attitudes, Values, and Personality        &  0.0\% &  7.4\% &  3.8\% &  61 \\
Family and Social Networks                &  7.9\% &  8.7\% &  7.1\% &  79 \\
Health and Care                           &  0.0\% &  6.7\% &  6.0\% &  60 \\
Integration, Migration, Transnation.      & 18.4\% & 18.8\% & 22.4\% & 185 \\
Housing, Equipment, Private HH Services   &  2.6\% & 15.1\% & 16.4\% & 141 \\
Time Use and Environmental Behaviour      &  0.0\% &  4.1\% &  5.5\% &  40 \\
\midrule
\textbf{Total}                            & \textbf{38} & \textbf{728} & \textbf{183} & \textbf{949} \\
\midrule
\emph{by scale type:} & & & & \\
\quad Nominal                             & 60.5\% & 48.8\% & 46.4\% & 463 \\
\quad Ordinal                             & 15.8\% & 25.5\% & 27.3\% & 242 \\
\quad Metric                              & 23.7\% & 25.7\% & 26.2\% & 244 \\
\bottomrule
\end{tabular}
\begin{tablenotes}
\small
\item \emph{Note.} Demographic Information = core demographic variables included as baseline context in all conditions. Questions for Persona Construction and Questions for Persona Evaluation were drawn randomly (80/20 split, seed$=$42) from the pool of 911 standard questions; any differences in distribution between the two splits occurred by chance. SOEP categories translated from the original German thematic areas (``Themen'').
\end{tablenotes}
\end{table}

%% file: 04_results.tex
\section{Results}
\label{sec:results}

We evaluated the $60$-cell construction-method grid (3 models $\times$ 5 information depths $\times$ 2 embedding methods $\times$ 2 reasoning modes; Section~\ref{sec:methodology}) on a held-out evaluation set of 183 questions (89 nominal, 94 ordinal/metric; per-cell coverage $n_q \approx 89$ after NA drops). 98.01\,\% of the 2{,}104{,}164 expected responses yielded valid scores; the remaining 1.99\,\% split into ambiguous ground truths (1.65\,\%), unparseable LLM outputs (0.32\,\%), and upstream pipeline errors (0.02\,\%). A per-model breakdown of data quality is provided in Web Appendix D.1. Sections~\ref{sec:results_depth}--\ref{sec:results_reasoning} and Web Appendix D report the detailed findings; Table~\ref{tab:construction_methods_grid_acc_corr} gives the joint per-cell accuracy and correlation grid.

\input{table_construction_methods_grid_acc_corr_20260525_223026}

\FloatBarrier

\subsection{Effects of the Information Depth}
\label{sec:results_depth}

\begin{figure}[htbp]
    \centering
    \includegraphics[width=0.49\linewidth]{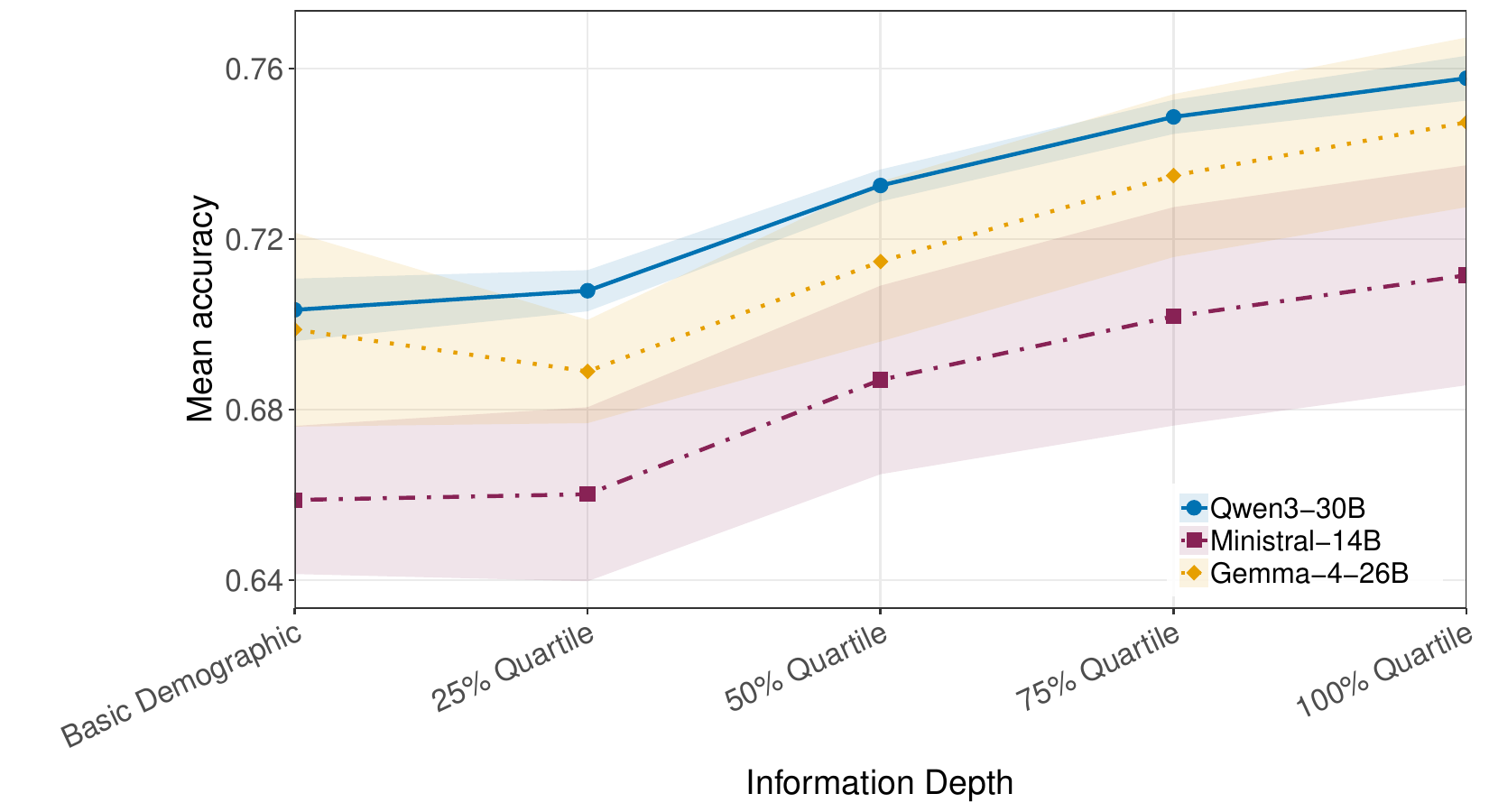}\hfill
    \includegraphics[width=0.49\linewidth]{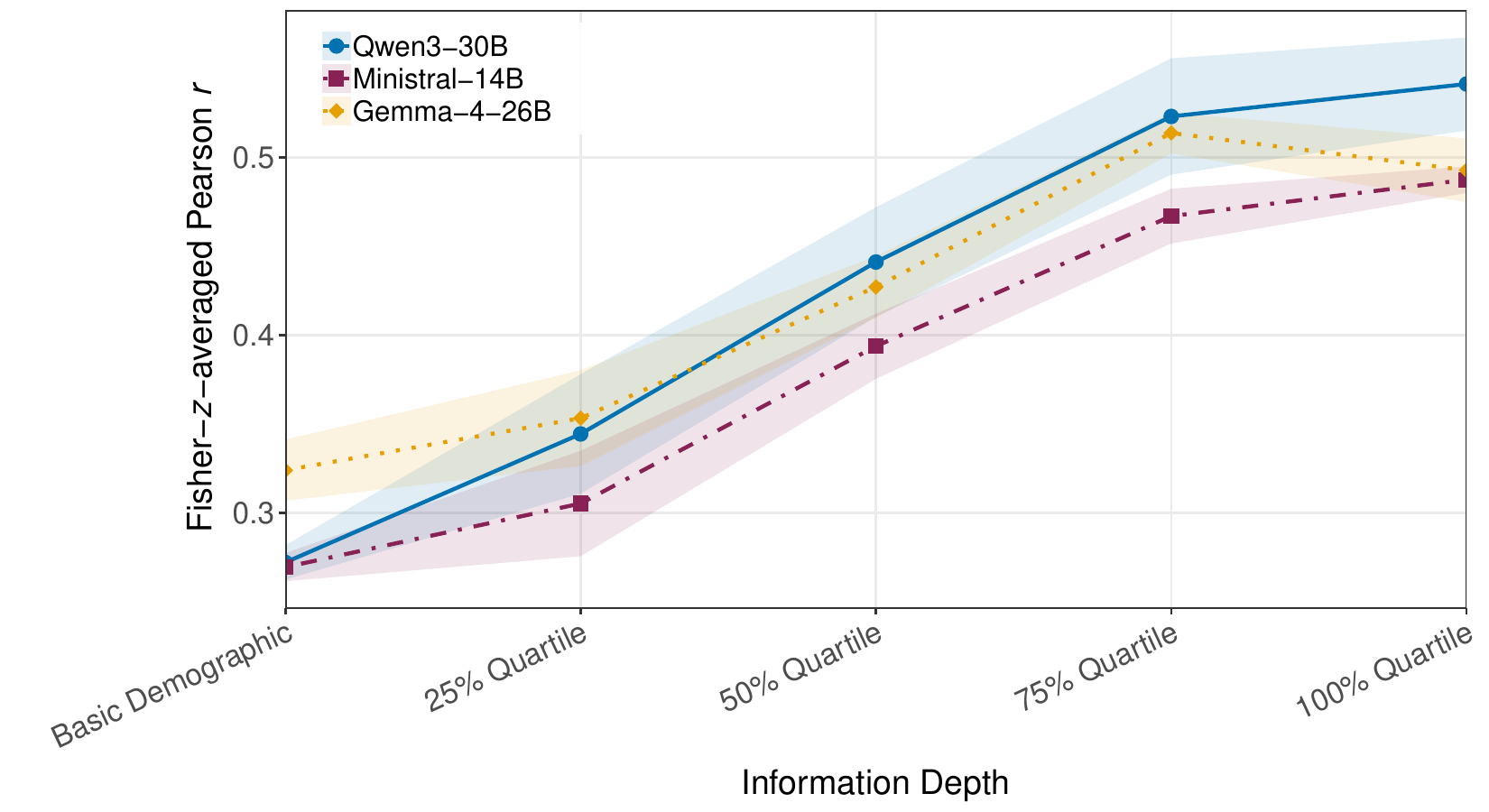}
    \caption{Accuracy (left) and Fisher-$z$ correlation (right) by information depth, with $\pm 1$ standard-error bands per model computed across the four construction cells (embedding $\times$ reasoning) at each depth.}
    \label{fig:depth_marginal_acc_corr}
\end{figure}

The marginal effect of cumulatively adding more persona context is summarized in Figure~\ref{fig:depth_marginal_acc_corr}, which plots accuracy and Fisher-$z$ correlation by depth. Both metrics increase monotonically across condition cells. Accuracy gained an average of $5.2$~pp from Basic Demographic to the 100\,\% Quartile, with per-model lifts of $+5.6$, $+5.1$, and $+6.8$~pp on the three Dialog Non-Thinking cells for Qwen 3, Gemma 4, and Ministral 3 respectively. Fisher-$z$ correlation gained an average of $21.9$~pp over the same range, a factor of four larger; Qwen 3 Dialog Non-Thinking moved from $r = 0.265$ to $0.504$ ($\Delta = +23.9$~pp), and its Thinking sibling reached the cross-grid maximum $r = 0.590$ at the 100\,\% Quartile ($\Delta = +29.1$~pp; Table~\ref{tab:construction_methods_grid_acc_corr}).
Correlation responded to persona content more sensitively than accuracy on every model that responded monotonically to depth, replicating the pattern that \textcite{peng2025digital} report on the Twin-2K-500 dataset \parencite{Toubia2025}, namely that rank-order correlation responds more strongly to added persona context than mean accuracy does, but at three to five times larger magnitude on our open-weights stack.

Across the cells, the bulk of the gain landed in the two middle transitions ($25 \to 50$\,\% and $50 \to 75$\,\%): together they accounted for $+4.3$ of the $+5.2$~pp accuracy lift and $+16.7$ of the $+21.9$~pp correlation lift. The opening transition (Basic $\to 25$\,\%) was effectively flat on accuracy ($-0.1$~pp), and the closing transition ($75 \to 100$\,\%) added only $+0.6$~pp of correlation.
The biggest single-cell accuracy jumps landed in the 25\,\% $\to$ 50\,\% transition (Gemma 4 Dialog Thinking $+5.0$~pp; Ministral 3 Dialog Non-Thinking $+3.6$~pp). The 75\,\% $\to$ 100\,\% transition delivered less correlation gain on average ($+0.6$~pp, with Gemma 4 Dialog Thinking actually losing $-4.0$~pp). This means that the context-length increase from the 75\,\% to the 100\,\% information depth did not pay off consistently on rank-ordering. The 75\,\% Quartile ($\approx 584$ of the 728 train items) captured most of the available accuracy signal and the bulk of the correlation signal.

\begin{figure}[htbp]
    \centering
    \includegraphics[width=\linewidth]{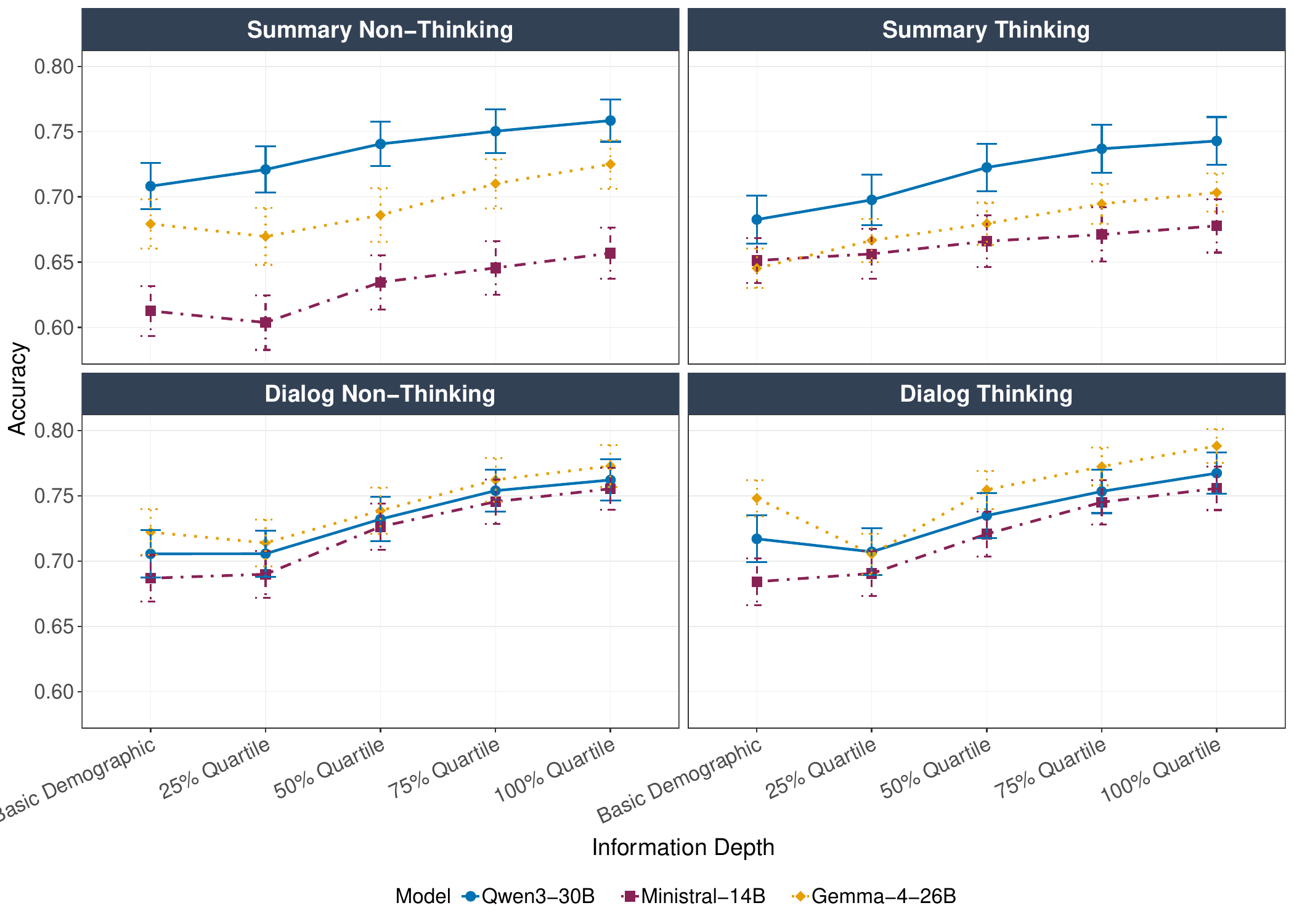}
    \caption{Per-model accuracy by information depth, with all four construction methods overlaid. }
    \label{fig:lineplot_by_depth_model_comparison_acc}
\end{figure}

The per-model accuracy curves (Figure~\ref{fig:lineplot_by_depth_model_comparison_acc}) show this concave pattern, with some model-specific heterogeneity across the construction methods.

\begin{figure}[htbp]
    \centering
    \includegraphics[width=0.8\linewidth]{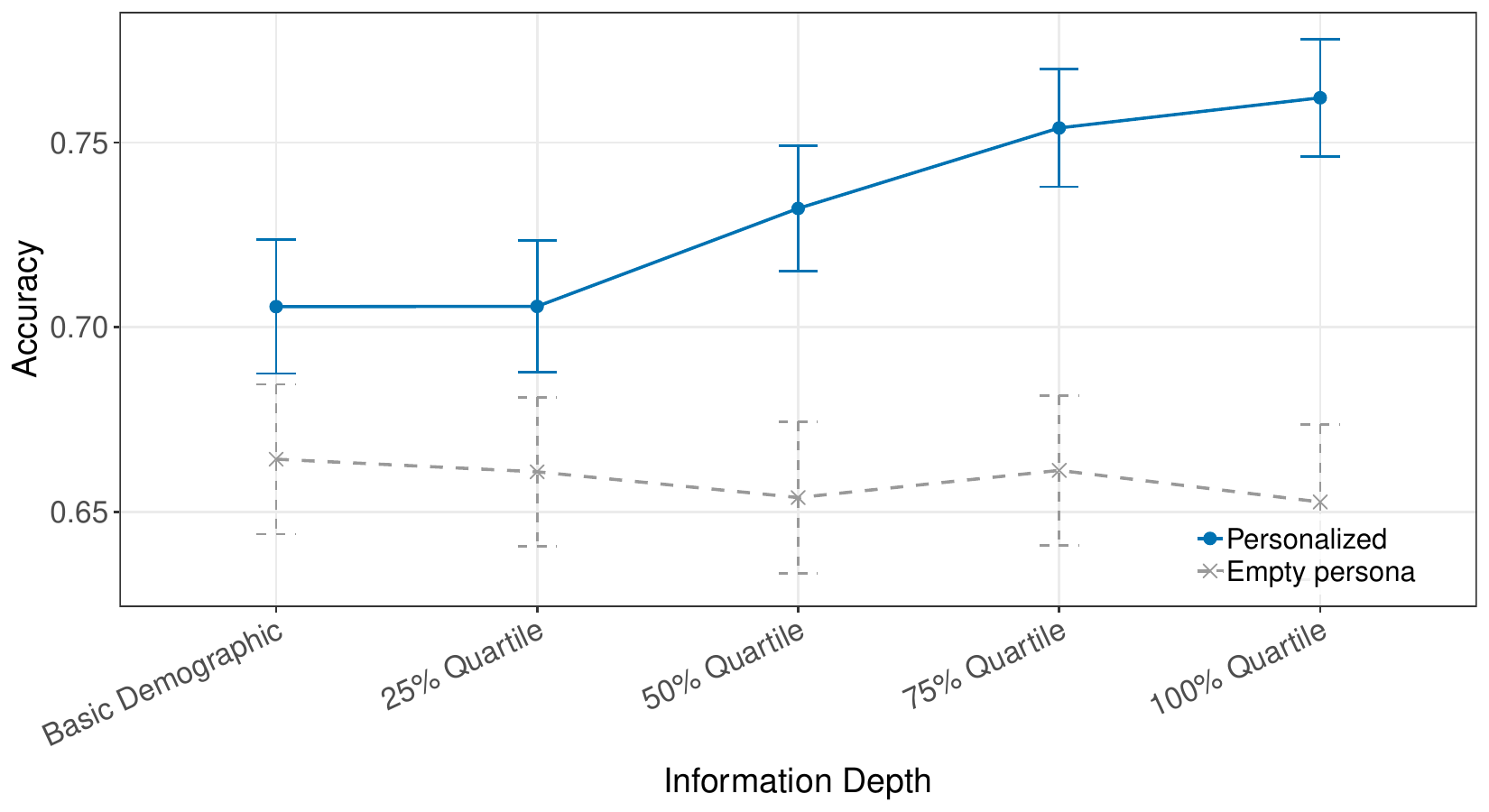}
    \caption{Empty-persona ablation. Personalized versus empty-persona accuracy by information depth on Qwen 3 Dialog Non-Thinking ($n=500$ participants, 183 held-out questions). Empty-persona accuracy is depth-flat at 0.65--0.66; the personalization delta widens from $+4.2$~pp at Basic Demographic to $+10.8$~pp at the 100\,\% Quartile.}
    \label{fig:persona_status_accuracy_main}
\end{figure}

To bound how much of the accuracy and correlation levels came from persona content rather than from instruction-following, we ran an empty-persona ablation where Qwen 3 Dialog Non-Thinking was re-evaluated at all five depths with the persona content replaced by an empty block (Figure~\ref{fig:persona_status_accuracy_main}, Web Appendix D.5). Empty-persona accuracy was flat, as expected, across information depths\footnote{This behavior is expected, since the model answered the same evaluation questions using the same context across all information depths. The variance is caused by the non-deterministic behavior of the model at a temperature level greater than 0.} at 0.65--0.66, while personalized accuracy climbed from 0.706 to 0.762, so the personalization delta widened from $+4.2$~pp at Basic Demographic to $+10.8$~pp at the 100\,\% Quartile, an order of magnitude larger than Peng's $+1.4$~pp on Twin-2K-500. Empty-persona correlation was structurally pinned at $r = 0$ at every depth (zero twin variance implies Pearson is coded as 0 per the \textcite{peng2025digital} convention), so the entire personalized correlation curve was attributable to persona content. The empty-persona correlation panel is reproduced in Web Appendix D.5.

We also found that the entropy-based ranking of persona-context questions had less effect than hypothesized. Cumulative random quartiles of equal size produced curves statistically indistinguishable from entropy-ranked quartiles at every depth on both metrics (paired-question sign test, $p > 0.10$ at every depth; aggregated paired test across the four ablation depths, $p \approx 0.32$; Web Appendix D.6). The increase in accuracy and correlation with depth was driven primarily by cumulatively adding more information, not by the within-pool prioritization of high-entropy items. Entropy is also not the only meaningful selection signal: per-item relevance to the evaluation question and per-question covariance structure are both plausible alternatives that we did not test here, and either could outperform random selection without being captured by raw response entropy.

\begin{figure}[htbp]
    \centering
    \includegraphics[width=0.82\linewidth]{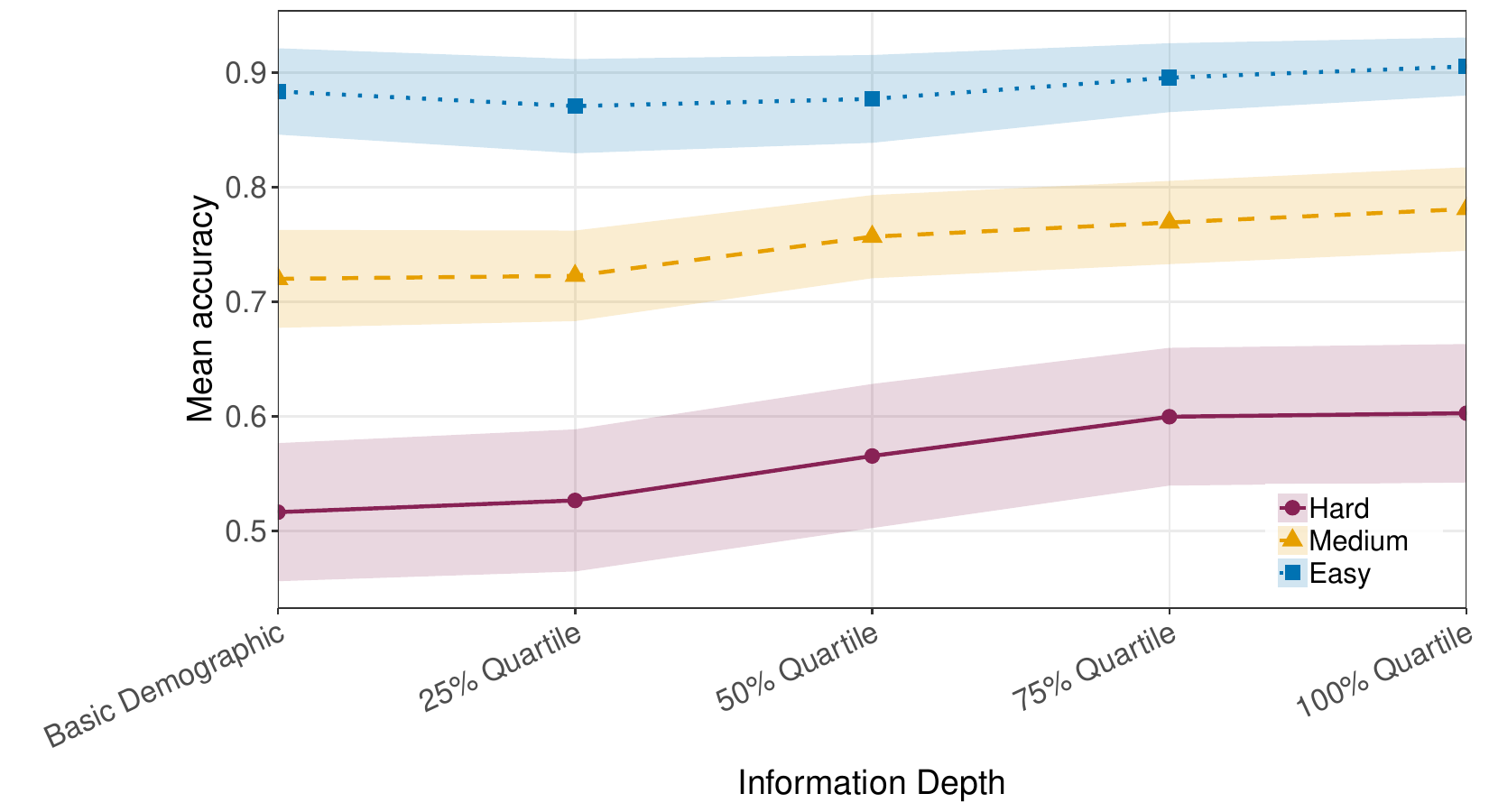}
    \caption{Personalized accuracy by information depth (Qwen 3 Dialog Non-Thinking, $n=500$ participants, 183 held-out items), split by difficulty tercile with 95\,\% confidence bands. Difficulty is defined by empty-persona accuracy at the 100\,\% Quartile. The hard tercile (red) absorbed $3.9\times$ the lift of the easy tercile (blue) over the depth range.}
    \label{fig:depth_by_difficulty}
\end{figure}

To better understand which questions drive the increase in accuracy and correlation, we split the held-out items into three difficulty terciles using empty-persona accuracy at the 100\,\% Quartile (cut-points 0.578 and 0.820; 59 hard / 58 medium / 58 easy items) and replotted the depth-accuracy curve within each tercile on the same Qwen 3 Dialog Non-Thinking run (Figure~\ref{fig:depth_by_difficulty}). Interestingly, the depth gain concentrated on the hard tercile, with the easy tercile near-saturated at every depth (88.4\,\%--90.6\,\% across all five depths). The corresponding deltas table is in Web Appendix D.10. Accuracy increased by $+8.6$~pp on hard items, $+6.1$~pp on medium items, and $+2.2$~pp on easy items. More information depth, in other words, helps most on items where the most common population response (the modal answer) is not the same as the participant's actual response, and barely helps on items where the modal response already coincides with the participant's answer.

Variance ratio shows a similar pattern to accuracy and correlation under cumulative information depth. Across all available construction cells the mean variance ratio rose from $0.98$ at Basic Demographic to $1.09$ at the 100\,\% Quartile ($+0.11$, a $+10.7\,\%$ relative increase). 

\FloatBarrier

\subsection{Effects of the Model Selection}
\label{sec:results_models}

\begin{figure}[htb]
    \centering
    \includegraphics[width=0.85\linewidth]{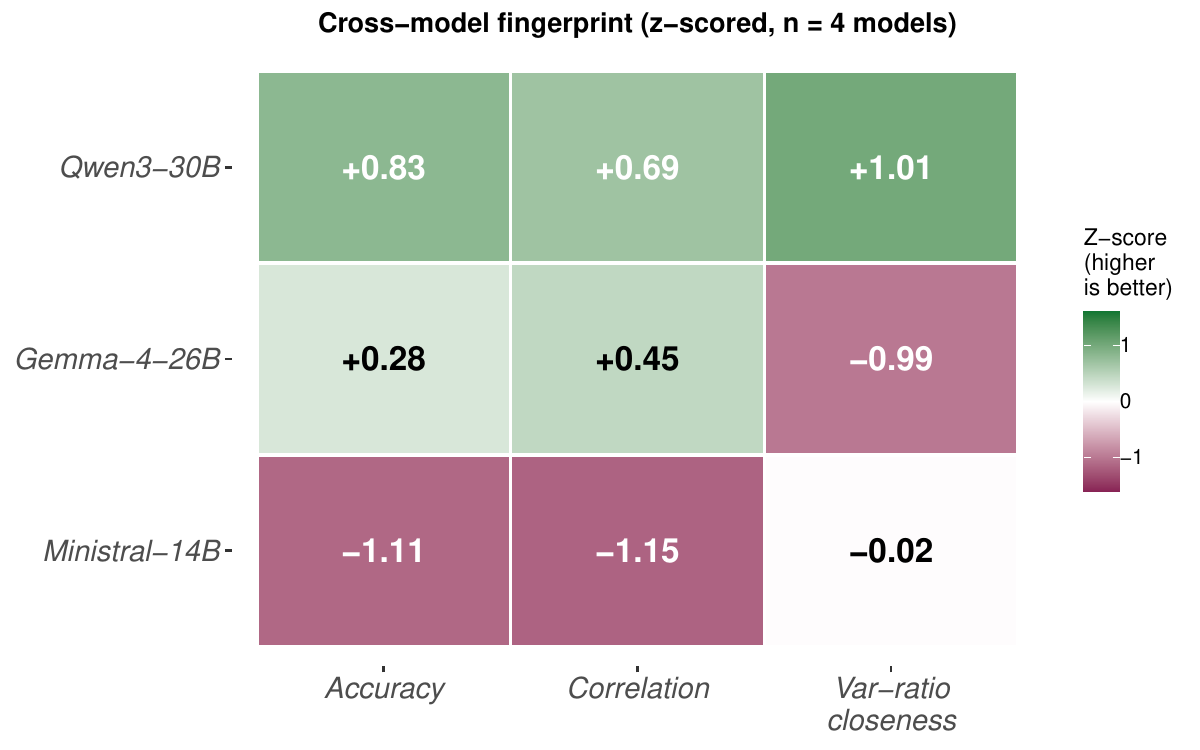}
    \caption{Z-scored composite fingerprint for the three models across the three twin-quality metrics. Each cell is the across-model z-score of the model's per-cell-averaged metric value (accuracy / Fisher-$z$ correlation / closeness to human variance, encoded as $-|\sigma_{\text{twin}}/\sigma_{\text{human}} - 1|$ so higher = closer to parity = better). Higher and greener = better on all three columns.}
    \label{fig:fingerprint_matrix}
\end{figure}

The three models separated cleanly when each construction cell's metric values were averaged into a per-model summary. The fingerprint matrix (Figure~\ref{fig:fingerprint_matrix}, Table~\ref{tab:fingerprint_matrix}) condenses each model into a $1 \times 3$ z-score row across the three metrics, averaged across the 20 construction cells per model. Qwen 3, on average, led every metric: accuracy ($0.730$), Fisher-$z$ correlation ($0.431$), and dispersion closeness ($\sigma_{\text{twin}}/\sigma_{\text{human}} = 0.99$), with composite row mean $+0.84$. Gemma 4 came in close on accuracy and correlation ($0.717$ and $0.426$) but is slightly under-dispersed (mean ratio $0.88$, composite $-0.09$). Ministral 3 trailed on accuracy ($0.684$) and correlation ($0.388$) and was mildly over-dispersed (mean ratio $1.07$, composite $-0.76$). The model ranking is not the same across embedding methods: on Persona Summary cells Qwen 3 leads at every information depth from the 25\,\% Quartile upward, while on Dialog cells Gemma 4 takes the cross-model top accuracy at all depths beyond the basic-demographic baseline, including the cross-grid maximum at 100\,\% Quartile Dialog Thinking ($0.788$). Ministral 3 trails on Persona at every depth and closes the gap to Gemma only on Dialog.

\begin{table}[htbp]
\caption{Per-model summary statistics and z-scored composite fingerprint, pooled across the 20 construction cells per model ($n{=}500$ participants; correlation and variance ratio restricted to ordinal/metric items). The z-score columns benchmark each model against the three-model panel; variance-ratio z-scores are computed on $-|\sigma_{\text{twin}}/\sigma_{\text{human}}-1|$ so that higher z = closer to human dispersion.}
\label{tab:fingerprint_matrix}
\centering
\input{fingerprint_matrix_20260525_222825}
\end{table}

Across the construction grid (Table~\ref{tab:construction_methods_grid_acc_corr}), every model scored its best accuracy in the Dialog $\times$ Thinking $\times$ 100\,\% Quartile cell, giving a consistent recipe for the accuracy-best configuration. Gemma 4's Dialog Thinking 100\,\% cell achieved the cross-grid maximum accuracy ($0.788$), Qwen 3's Dialog Thinking 100\,\% cell achieved the cross-grid maximum correlation ($r = 0.590$), and Ministral 3 benefited disproportionately from the Dialog embedding ($+9.8$~pp at 100\,\% Quartile Non-Thinking, the largest single-cell embedding effect in the grid; see Section~\ref{sec:results_embedding}).

\begin{figure}[htb]
    \centering
    \includegraphics[width=\linewidth]{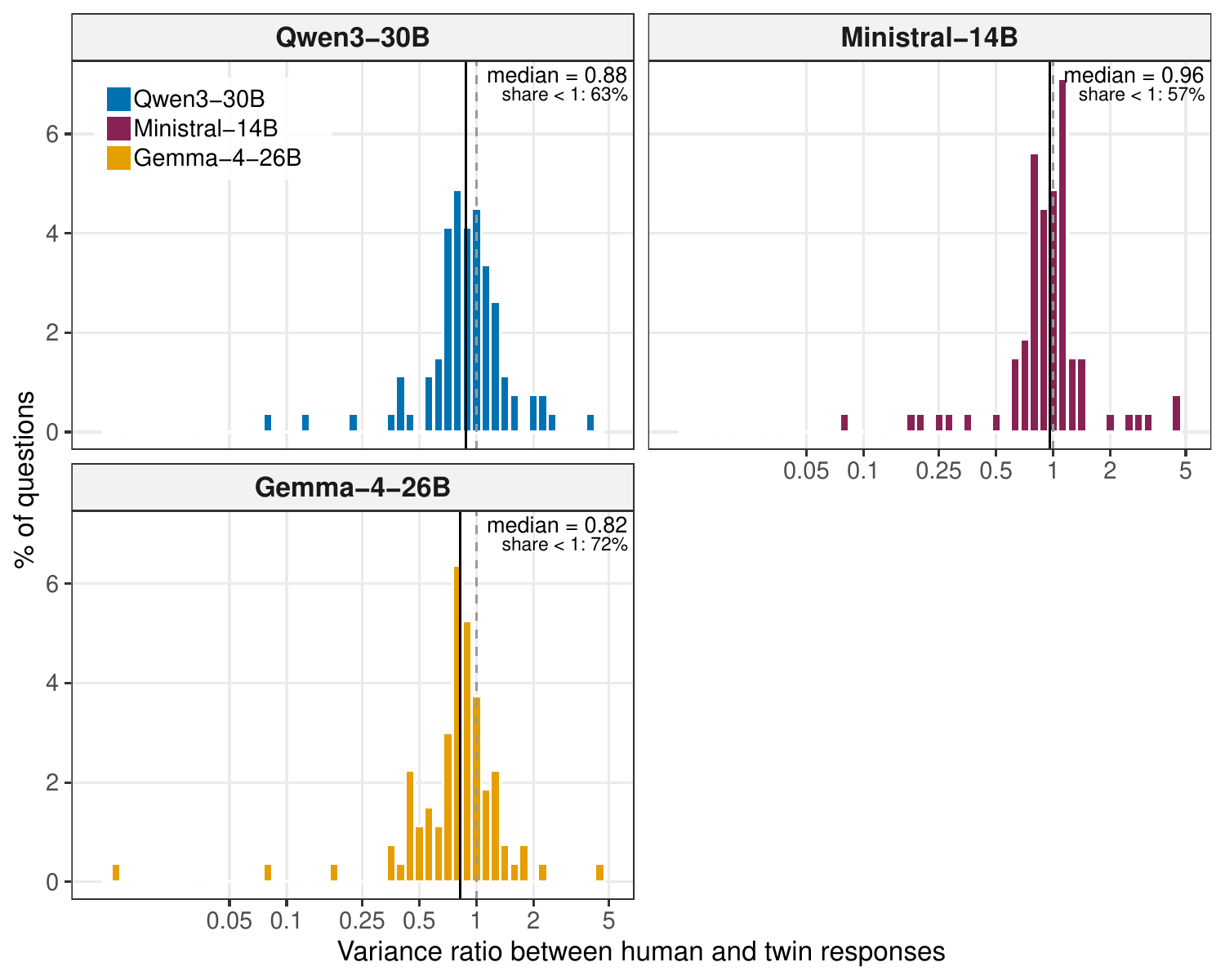}
    \caption{Per-question variance-ratio distribution $\sigma_{\text{twin}}/\sigma_{\text{human}}$ by model on a log $x$-axis. Each per-(model, question) value is the mean across the 20 construction cells per model of the per-cell winsorised SD ratio on that question; per-cell ratios are capped at~5 before averaging. Restricted to ordinal- and metric-scale items. Dashed grey line at ratio${=}1$; solid black line at the per-model median.}
    \label{fig:per_question_histogram_variance_ratio_main}
\end{figure}

The per-question variance-ratio distributions (Figure~\ref{fig:per_question_histogram_variance_ratio_main}) are well-shaped for all three models, with per-model medians close to 1 (Qwen $\approx 0.99$, Ministral $\approx 1.07$) or modestly below (Gemma $\approx 0.88$) and a reasonable spread.

\subsection{Effects of the Embedding Method}
\label{sec:results_embedding}
\begin{figure}[htbp]
    \centering
    \includegraphics[width=0.32\linewidth]{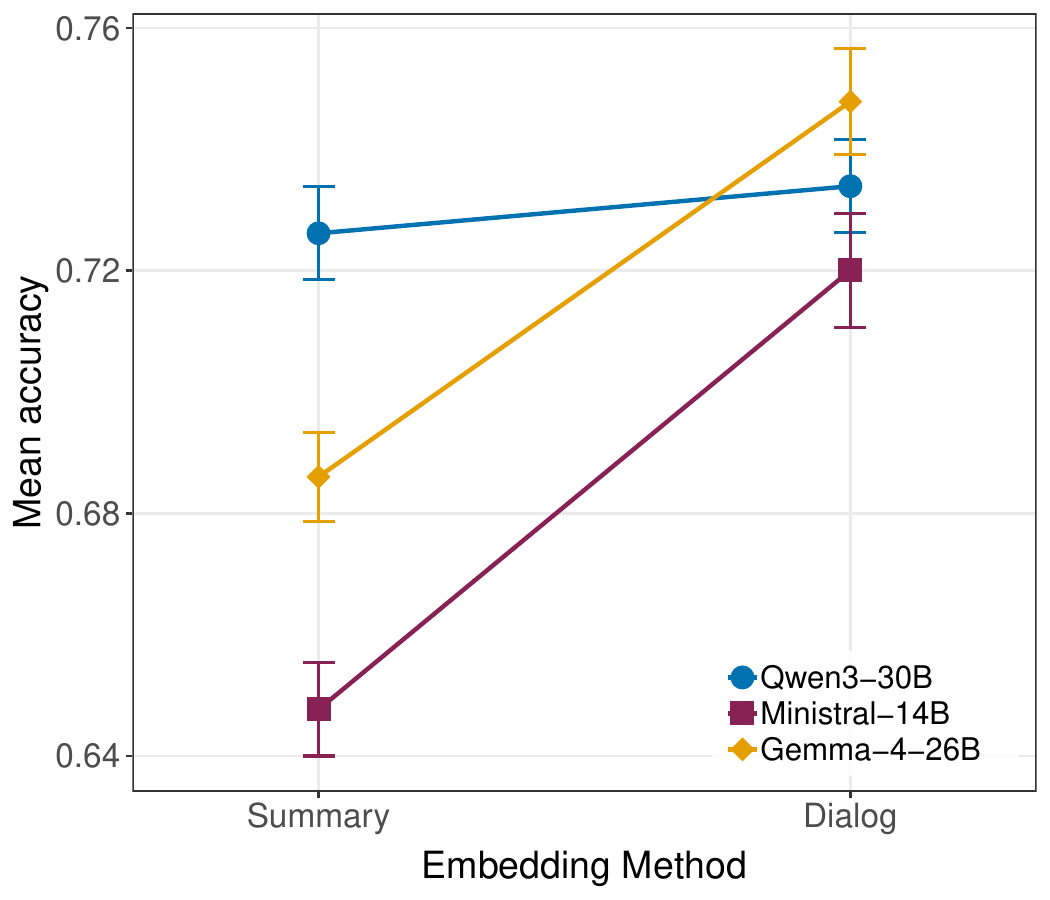}\hfill
    \includegraphics[width=0.32\linewidth]{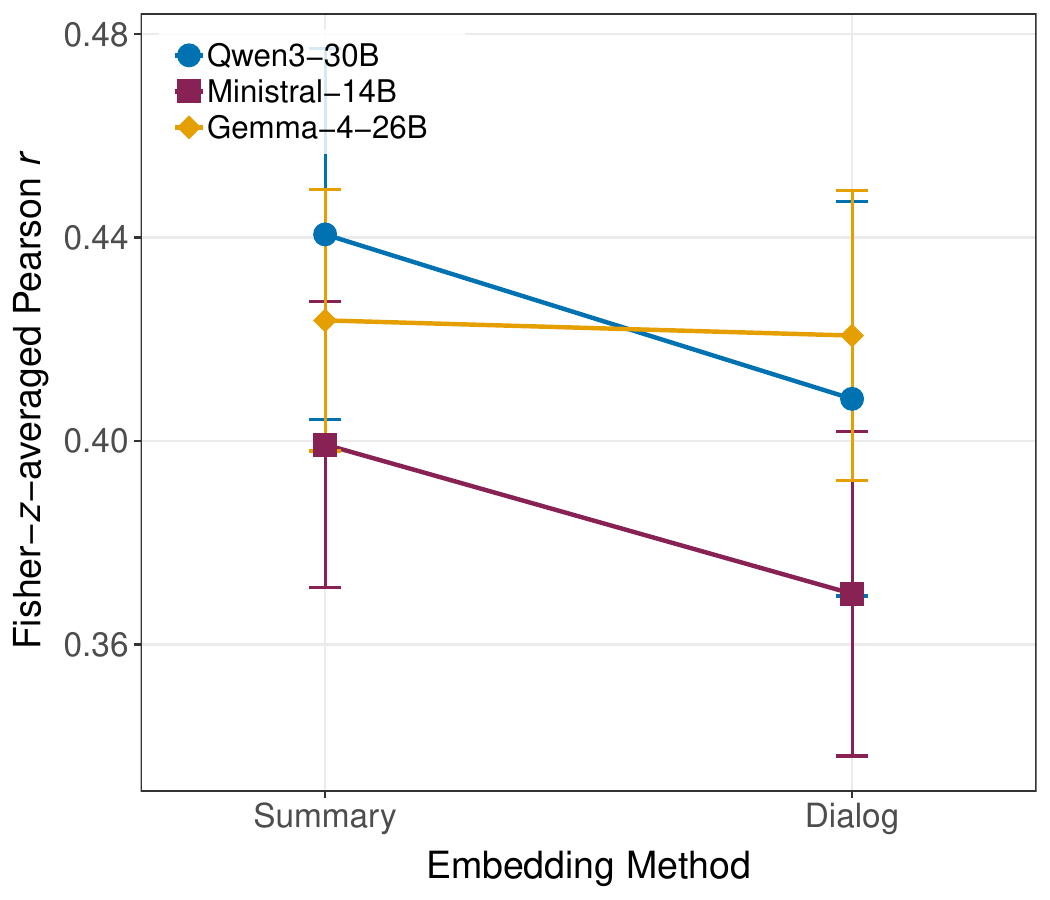}\hfill
    \includegraphics[width=0.32\linewidth]{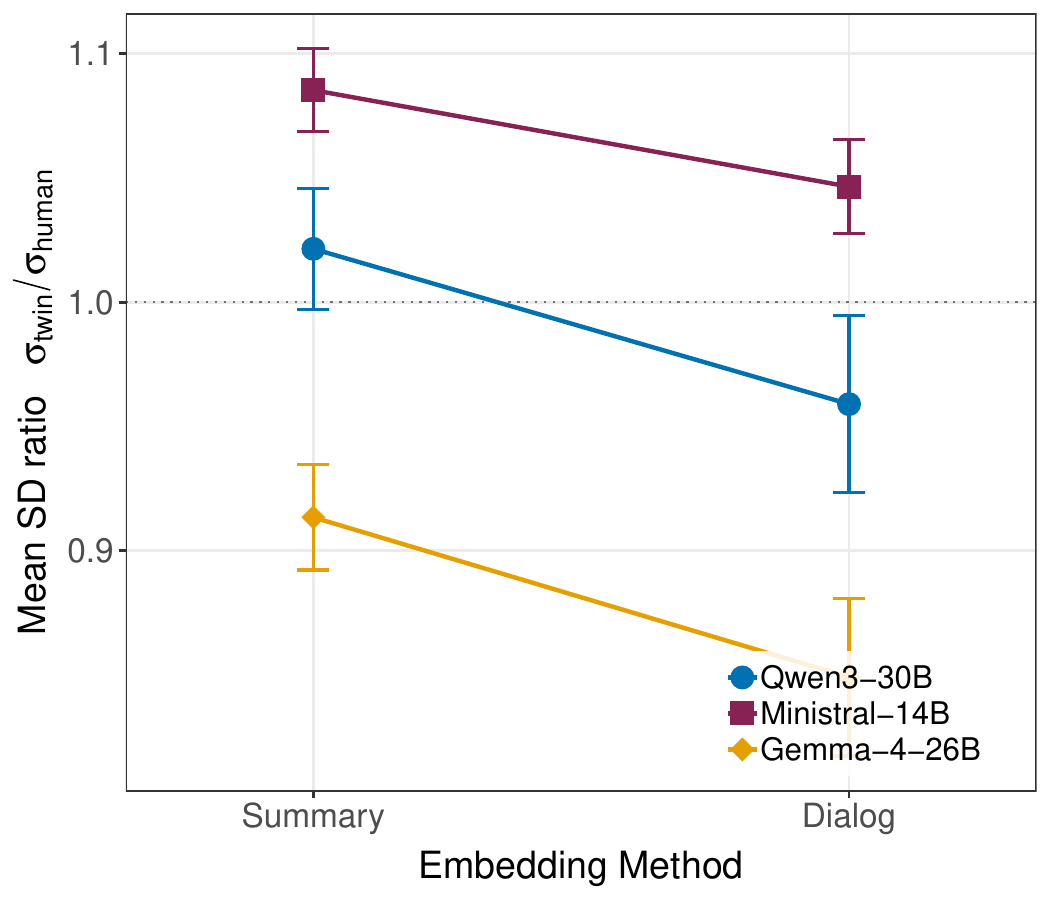}
    \caption{Persona summary (left endpoint) versus Dialog input (right endpoint) per model, pooled across reasoning modes and information depths. Left: accuracy. Center: Fisher-$z$ correlation. Right: variance ratio.}
    \label{fig:embedding_dumbbells}
\end{figure}

We compared the two embedding strategies, the Chain-of-Density (CoD) Persona summary and the Dialog input, on the same construction grid. Figure~\ref{fig:embedding_dumbbells} reports the marginal effect of switching from Persona to Dialog, pooled across reasoning modes and information depths, on each of the three twin-quality metrics.

Accuracy increased for all three models when switching from Persona to Dialog. The mean gain across construction cells was $+6.2$ percentage points for Gemma 4 (paired $t$-test $p = 8.3 \times 10^{-6}$) and $+7.2$ percentage points for Ministral 3 ($p = 6.1 \times 10^{-6}$); Qwen 3's gain was much smaller at $+0.8$ percentage points ($p = 0.14$). Fisher-$z$ correlation barely moved for any model (all $|\Delta r| \leq 0.03$, $p \geq 0.07$). The variance ratio stayed close to parity for Ministral 3 ($1.09 \to 1.05$, $p = 0.08$) and Qwen 3 ($1.02 \to 0.96$, $p = 0.22$); for Gemma 4 the ratio moved further from parity ($0.91 \to 0.85$, $p = 0.026$). Per-cell numbers underlying these aggregates are in Web Appendix D.2 (accuracy), Web Appendix D.3 (correlation), and Web Appendix D.4 (variance ratio).

Restricting attention to the 100\,\% Quartile cells, switching from Persona to Dialog raised accuracy in every (model $\times$ reasoning) cell (Web Appendix D.2). The largest single-cell improvements were Ministral 3 Non-Thinking ($+9.8$~pp), Gemma 4 Thinking ($+8.5$~pp), and Ministral 3 Thinking ($+7.8$~pp); Qwen 3, by contrast, picked up at most $+2.4$~pp from the Dialog embedding at 100\,\% depth.

The accuracy gain comes at a measurable token cost. At the 100\,\% Quartile, the Persona summary averages 2{,}761 words / 5{,}354 tokens per participant; the Dialog input averages 7{,}074 words / 14{,}863 tokens, roughly $2.6\times$ the words and $2.8\times$ the tokens (sampled $n=100$ participants; \texttt{tiktoken cl100k\_base}). The CoD summary's token efficiency comes from paraphrasing scale items into integrated prose, but that compression collapses information, like Likert labels, into approximate descriptions. A participant's response of ``Does not apply at all'' (lowest of six points) to the question ``How much do you agree with the statement that you react annoyed when others take your attention away?'' (translated from German) is rendered in the CoD summary as ``She does not react annoyed when others take her attention away.'' (translated from German). Dialog preserves the exact information, while persona summaries collapse it to a binary negation.

The trade-off is therefore one of lossy compression: the Dialog input keeps every response from past surveys exactly as stated in context, while the Persona summary saves roughly $60\,\%$ of the token cost (and a proportional share of latency and prompt-cache footprint) at the price of precision.

\FloatBarrier

\subsection{Effects of the Reasoning Mode}
\label{sec:results_reasoning}
\begin{figure}[htbp]
    \centering
    \includegraphics[width=0.49\linewidth]{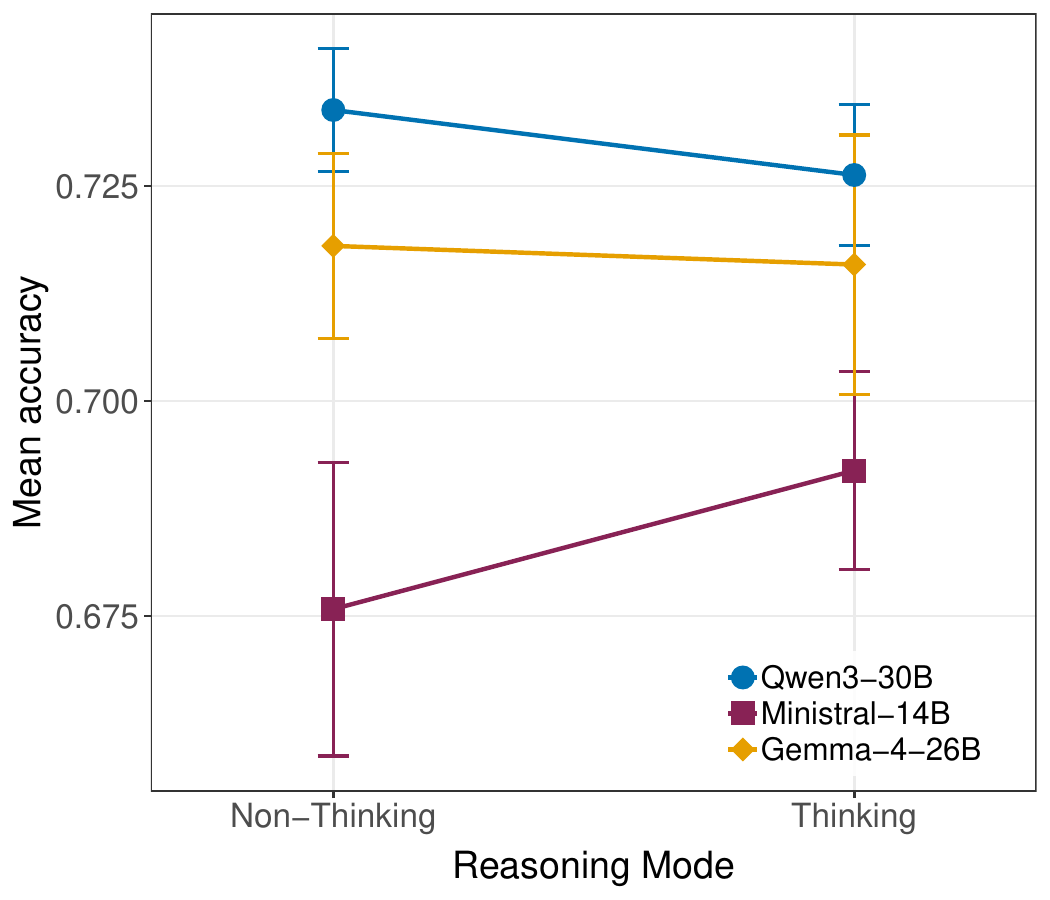}\hfill
    \includegraphics[width=0.49\linewidth]{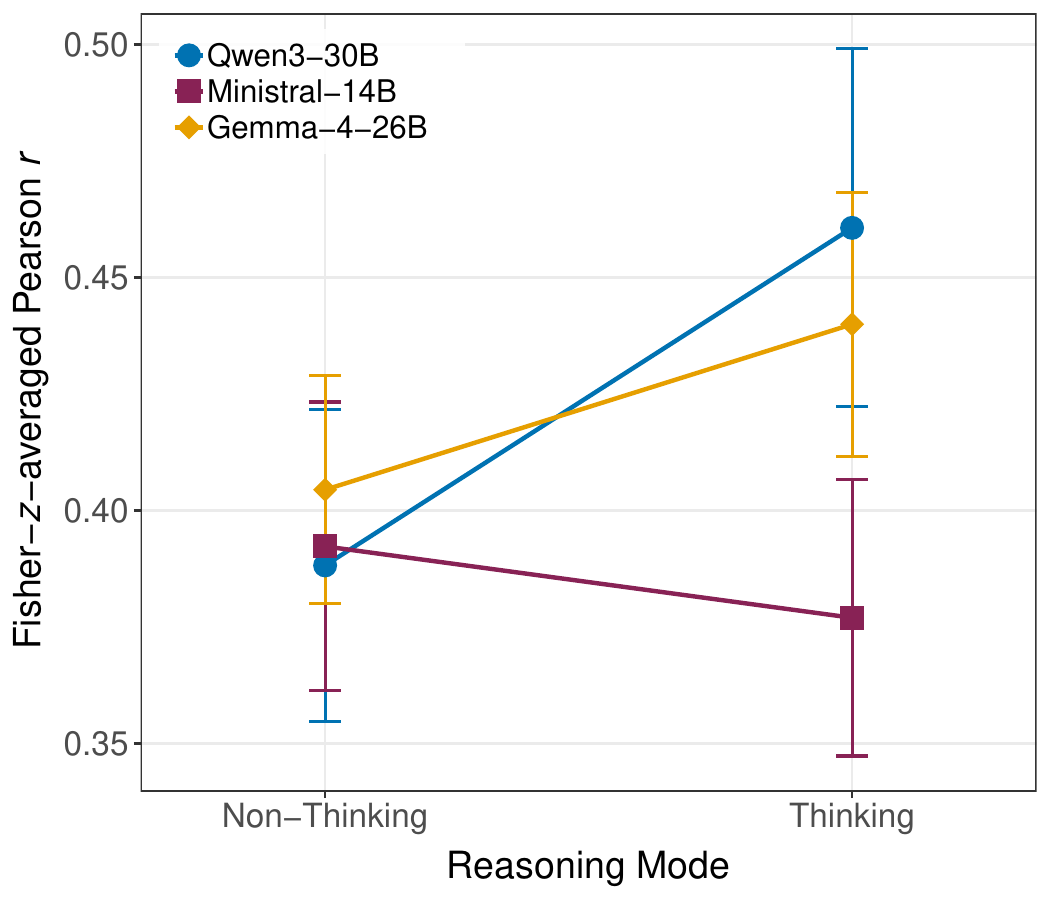}
    \caption{Non-Thinking (left endpoint) versus Thinking (right endpoint) per model, pooled across embedding methods and information depths. Left: accuracy. Right: Fisher-$z$ correlation.}
    \label{fig:reasoning_dumbbells}
\end{figure}

We compared Non-Thinking and Thinking by pairing the matched (embedding $\times$ information depth) cells within each model. Figure~\ref{fig:reasoning_dumbbells} reports the marginal accuracy and correlation effect; the matched variance-ratio dumbbell is in Web Appendix D.8.

Accuracy barely moved between reasoning modes for any model (all $|\Delta_{\text{acc}}| \leq 2$ percentage points across cells; only Ministral 3's $+1.6$ percentage-point shift reached a paired-$t$ $p$-value below 0.05). Fisher-$z$ correlation rose under Thinking for Qwen 3 ($\Delta r = +0.073$, paired $t$-test $p = 3 \times 10^{-4}$) and Gemma 4 ($\Delta r = +0.036$, $p = 0.014$), while Ministral 3 was effectively flat ($\Delta r = -0.015$, $p = 0.22$). The variance ratio moved toward parity under Thinking, most visibly on Gemma 4 ($0.83 \to 0.93$, $p = 0.004$).

Restricting attention to the 100\,\% Quartile cells, every (model $\times$ embedding) cell showed a directional correlation increase under Thinking, with Qwen 3's Dialog 100\,\% Thinking cell taking the cross-grid maximum at $r = 0.590$ (up from $0.504$ under Non-Thinking). Thinking helped most where the base configuration was already well-aligned. The largest $\Delta r$ at the 100\,\% Quartile landed on Qwen 3 Summary ($+0.094$), Qwen 3 Dialog ($+0.086$), and Gemma 4 Summary ($+0.086$); Ministral 3's gains were an order of magnitude smaller (Summary $+0.022$, Dialog $+0.021$).

The correlation gain came at a substantial cost in inference time. Thinking inference was substantially slower per prediction than Non-Thinking on our hardware, with per-model ratios spanning roughly an order of magnitude: Qwen 3 $21\times$, Gemma 4 $5\times$, and Ministral 3 $3\times$. The cost reflects both the additional reasoning-trace tokens and the wider generation budget configured for Thinking runs.

%% file: table_construction_methods_grid_acc_corr_20260525_223026.tex
\begin{table}[htbp]
\caption{Accuracy and correlation by construction method and model ($n{=}500$ participants).}
\label{tab:construction_methods_grid_acc_corr}
\centering
\begin{tabular}{l cc cc}
\toprule
 & \multicolumn{2}{c}{\textbf{Persona Description}} & \multicolumn{2}{c}{\textbf{Dialog Input}} \\
\cmidrule(lr){2-3} \cmidrule(lr){4-5}
\textbf{Information Depth} & Non-Thinking & Thinking & Non-Thinking & Thinking \\
 & \textit{acc$\uparrow$ / corr$\uparrow$} & \textit{acc$\uparrow$ / corr$\uparrow$} & \textit{acc$\uparrow$ / corr$\uparrow$} & \textit{acc$\uparrow$ / corr$\uparrow$} \\
\midrule
\multicolumn{5}{l}{\textcolor[HTML]{0072B2}{\rule[0.05em]{0.7em}{0.7em}}~\textit{Qwen3-30B}} \\
Basic Demographic & 0.708 / 0.253 & 0.683 / 0.272 & 0.706 / 0.265 & \textbf{0.717} / \textbf{0.299} \\
25\% Quartile & \underline{\textbf{0.721}} / 0.369 & 0.698 / \textbf{0.403} & 0.706 / 0.248 & 0.707 / 0.359 \\
50\% Quartile & \textbf{0.741} / 0.456 & 0.723 / \underline{\textbf{0.490}} & 0.732 / 0.351 & 0.735 / 0.468 \\
75\% Quartile & 0.750 / 0.514 & 0.737 / \underline{\textbf{0.579}} & \textbf{0.754} / 0.434 & 0.753 / 0.565 \\
100\% Quartile & 0.759 / 0.489 & 0.743 / 0.582 & 0.762 / 0.504 & \textbf{0.767} / \underline{\textbf{0.590}} \\
\midrule
\multicolumn{5}{l}{\textcolor[HTML]{E69F00}{\rule[0.05em]{0.7em}{0.7em}}~\textit{Gemma-4-26B}} \\
Basic Demographic & 0.679 / 0.329 & 0.645 / 0.278 & 0.722 / 0.327 & \underline{\textbf{0.748}} / \underline{\textbf{0.362}} \\
25\% Quartile & 0.670 / 0.368 & 0.667 / \underline{\textbf{0.418}} & \textbf{0.714} / 0.290 & 0.705 / 0.337 \\
50\% Quartile & 0.686 / 0.405 & 0.680 / \textbf{0.459} & 0.739 / 0.390 & \underline{\textbf{0.755}} / 0.454 \\
75\% Quartile & 0.710 / 0.486 & 0.695 / 0.515 & 0.762 / 0.512 & \underline{\textbf{0.772}} / \textbf{0.542} \\
100\% Quartile & 0.725 / 0.446 & 0.703 / \textbf{0.532} & 0.773 / 0.491 & \underline{\textbf{0.788}} / 0.502 \\
\midrule
\multicolumn{5}{l}{\textcolor[HTML]{882255}{\rule[0.05em]{0.7em}{0.7em}}~\textit{Ministral-14B}} \\
Basic Demographic & 0.613 / 0.250 & 0.651 / \textbf{0.286} & \textbf{0.687} / 0.277 & 0.684 / 0.265 \\
25\% Quartile & 0.604 / 0.354 & 0.656 / \textbf{0.358} & 0.690 / 0.268 & \textbf{0.690} / 0.242 \\
50\% Quartile & 0.635 / \textbf{0.437} & 0.666 / 0.385 & \textbf{0.726} / 0.404 & 0.721 / 0.349 \\
75\% Quartile & 0.646 / \textbf{0.504} & 0.671 / 0.430 & \textbf{0.746} / 0.476 & 0.745 / 0.458 \\
100\% Quartile & 0.657 / 0.484 & 0.678 / \textbf{0.506} & 0.755 / 0.470 & \textbf{0.756} / 0.491 \\
\bottomrule
\end{tabular}
\begin{tablenotes}
\small
\item \emph{Note.} \textbf{Bold} marks the best value within each (model, depth) row; \underline{underline} marks the best value across all models within each depth. Both are applied independently for accuracy and correlation.
\end{tablenotes}
\end{table}

%% file: fingerprint_matrix_20260525_222825.tex
\begin{tabular}{l rrr rrr r}
\toprule
\textbf{Model} & \textbf{Acc.} & \textbf{Corr.} & $\sigma_{\text{twin}}/\sigma_{\text{human}}$ & z(Acc) & z(Corr) & z(VR-close) & z(row mean) \\
\midrule
Qwen3-30B & 0.730 & 0.431 & 0.990 & +0.83 & +0.69 & +1.01 & \textbf{+0.84} \\
Gemma-4-26B & 0.717 & 0.426 & 0.881 & +0.28 & +0.45 & -0.99 & \textbf{-0.09} \\
Ministral-14B & 0.684 & 0.388 & 1.066 & -1.11 & -1.15 & -0.02 & \textbf{-0.76} \\
\bottomrule
\end{tabular}

%% file: 05_discussion.tex
\section{Discussion and Conclusion}
\label{sec:discussion}

Four patterns emerge from the 60-cell construction grid. Accuracy and rank-order correlation both rise with information depth, with diminishing returns past the 75 percent entropy quartile. The 100 percent quartile contains the best-performing cells overall and represents the achievable ceiling on the panel data we have. The 75 percent quartile, in contrast, captures the bulk of that gain at a moderate data-acquisition and prompt-token cost, and is therefore the cost-efficient Pareto point firms should target when item volume is a constraint. The depth gain is also non-uniform across items: hard items, defined by low empty-persona accuracy, gain $+8.6$ percentage points on accuracy from basic demographics to the 100 percent quartile, while easy items gain only $+2.2$ percentage points. More persona context, in other words, helps most where the most common population response does not match the participant's actual answer. Switching the embedding from a Chain-of-Density persona summary to a raw dialog history of past survey responses raises accuracy in every model-by-reasoning cell at the 100 percent depth, while an explicit thinking mode lifts Fisher-$z$ correlation by an average of $+0.038$ across models while leaving accuracy roughly unchanged. A plausible interpretation is that thinking allocates extra inference compute to weighing competing persona cues against each other, which sharpens between-respondent contrasts without meaningfully improving mean accuracy across the population.

Within the open-weights stack, Qwen 3 leads on average across all three metrics (accuracy 0.730, Fisher-$z$ correlation 0.431, dispersion ratio 0.99), Gemma 4 attains the single best accuracy cell at 0.788, and Qwen 3 attains the highest rank-order correlation at $r = 0.590$, both in the dialog-thinking-100 percent cell.

Our headline accuracy and rank-order correlation sit at the same order of magnitude as recent findings in the digital-twin literature, despite our reliance on a pre-existing panel rather than a purpose-built instrument. Best-cell accuracy of 0.788 on the SOEP held-out evaluation set sits above the 0.748 full-persona accuracy reported by \textcite{peng2025digital} on Twin-2K-500 and above the 0.72 absolute reported by \textcite{Toubia2025}, while the 0.83 to 0.86 reported by \textcite{park2026llmagentsgroundedselfreports} is on a normalized scale that benchmarks twin agreement against participants' own two-week test-retest consistency and is therefore comparable in interpretation rather than in level. Best-cell rank-order correlation of $r = 0.590$ is roughly three times the per-outcome $r = 0.197$ that \textcite{peng2025digital} report on the same construct. The most plausible reading of the rank-order gap is that SOEP carries more individuating signal per added item than the Twin-2K-500 battery, whose personality and cognitive items tend to shrink toward base-model priors \parencite{peng2025digital}.

The empty-persona ablation tells the same story: the personalization delta widens with depth to roughly an order of magnitude larger than the $+1.4$ percentage-point delta reported by \textcite{peng2025digital}, consistent with the SOEP panel carrying more usable personalization signal than the Twin-2K-500 dataset. A random-quartile ablation produced curves indistinguishable from entropy-ranked quartiles (aggregated paired test $p \approx 0.32$), so it is the total volume of items in the prompt that drives twin quality, not the within-pool rule by which items are selected.

The distortion family cataloged by \textcite{peng2025digital} does not reproduce uniformly. Insufficient individuation shows up only on Gemma 4, whose mean variance ratio of 0.88 sits below human dispersion and which thinking mode partially corrects (0.83 to 0.93). Qwen 3 has a near-parity mean variance ratio of 0.99, and Ministral 3 is mildly over-dispersed at 1.07. The depth-by-difficulty split likewise indicates that easy items are saturated by demographics alone, so demographic shortcutting is concentrated on a specific subset of items rather than being a universal property of the twins.

Detailed individual-level digital twins, as we, \textcite{peng2025digital}, and \textcite{park2026llmagentsgroundedselfreports} build them, have a structural advantage over the coarse persona-bot approach used by most marketing twin work \parencite{brand2023using,li2024frontiers,wang2026large,goli2024frontiers}: they can be aggregated post hoc to any segment a researcher cares to define, including segments not anticipated at construction time, while a persona-bot designed to represent a particular demographic segment cannot be decomposed back into individuals. Combined with our results, where best-cell accuracy is comparable to bespoke-data twins and rank-order correlation is roughly three times the reported per-outcome level on Twin-2K-500, this means that post-hoc segmentation and segment-versus-segment comparisons built on these twins should reproduce both group-level means and within-group heterogeneity better than aggregations built from coarse persona bots.

For firms that hold heterogeneous panel data, a panel of roughly 1{,}000 question-answer pairs per customer is sufficient material to build twins that reach the same order of magnitude on accuracy as twins built on purpose-collected interviews or surveys. The construction grid maps a usable cost-quality frontier. For accuracy-first applications, the dialog-thinking-100 percent cell on Gemma 4 is the recommended recipe. For cost-first applications with acceptable accuracy, a persona-non-thinking-75 percent cell captures the bulk of the available accuracy gain at substantially lower compute. On input tokens, the Chain-of-Density persona summary at the 100 percent quartile uses roughly 36 percent of the dialog-condition input budget (5{,}354 versus 14{,}863 tokens per participant), so switching from Dialog to Persona saves about two-thirds of the prompt-token bill. On output tokens, the non-thinking mode avoids the reasoning-trace generation that lifts thinking-mode inference time by roughly $3\times$ (Ministral 3), $5\times$ (Gemma 4), and $21\times$ (Qwen 3) over direct prompting. For applications that prize rank-ordering of respondents, such as segmentation or scoring, the dialog-thinking-100 percent cell on Qwen 3 is preferred.

The depth-flat empty-persona baseline of 0.65 to 0.66 has an instructive interpretation rather than being a nuisance. Without any persona content, the LLM still answers a large share of test items correctly because many SOEP items have a modal answer that the model can reach by giving the socially desirable response, the most common response in the population, or the response a base model has internalized from pretraining. The baseline is therefore the right yardstick for separating instruction-following accuracy from the marginal value of individual persona content, and empty-persona depth ablations should become standard hygiene in twin papers, since otherwise the contribution of individual context to reported accuracy is uninterpretable.

\subsection{Limitations and Future Research}
\label{sec:limitations_future}

Several limitations bound the present results and motivate clear extensions. The SOEP is a German general-population panel, and brand and choice items typical of consumer marketing are not directly comparable. Future research should explore whether our findings can be reproduced using a US- or UK-panel on commercially relevant items. We evaluate three open-weights models that dominate the single-GPU open-source frontier in May 2026, but we do not benchmark against closed-source frontier models because of data-license constraints. Finally, we cover insufficient individuation, demographic shortcutting, and contamination via empty persona, but not ideological drift \parencite{lyman2025balancing,motoki2024more,rozado2024political} or persona-induced reasoning bias and diversity collapse \parencite{gupta2024bias,park2024diminished}, all of which warrant dedicated follow-up work.

\subsection{Conclusion}
\label{sec:conclusion}

Detailed individual-level digital twins do not require bespoke conditioning data. The pre-existing heterogeneous panel data that firms already accumulate through CRM systems, loyalty programs, and repeat surveys is sufficient material to build twins that reach the same order of magnitude on accuracy and rank-order correlation as published twins built on purpose-collected interviews or surveys. Within that data, the construction-method search space collapses into a small set of usable trade-offs: the dialog-thinking-100 percent recipe yields the strongest twins we observed, the persona-non-thinking-75 percent recipe yields a cost-efficient near-substitute, and the choice of model depends on whether the use case prizes accuracy, rank-order correlation, or dispersion fidelity. Combined, these findings move the question from ``can large firms with bespoke survey budgets build twins?'' to ``can any firm with a heterogeneous panel of customer responses build twins?''. The latter is now answered in the affirmative, and the path toward democratising consumer insights through LLM-based digital twin research is open.

\FloatBarrier